\documentclass[showpacs,preprintnumbers,amsmath,amssymb,twocolumn,superscriptaddress,prb]{revtex4}
\usepackage{times}
\usepackage{graphicx}
\usepackage{amsfonts}
\usepackage{amsmath, amsthm, amssymb}
\usepackage{dsfont}

\newcommand{\vk}{\mathbf{k}} 
\newcommand{\vp}{\mathbf{p}} 
\newcommand{\vecr}{\mathbf{r}} 
\newcommand{\vV}{\mathbf{V}} 
\newcommand{\vg}{\mathbf{g}} 
\newcommand{\e}[1]{\mathrm{e}^{#1}}
\newcommand{\cop}{\hat{c}} 
\newcommand{\A}{A_{\vk}} 

\newcommand{\ie}{\textit{i.e. }}
\newcommand{\eg}{\textit{e.g. }}
\newcommand{\etal}{\emph{et al.}}
\def\i{\mathrm{i}}

\begin{document}
\title[Quantum transport in noncentrosymmetric superconductors and thermodynamics of ferromagnetic superconductors]{Quantum transport in noncentrosymmetric superconductors and thermodynamics of ferromagnetic superconductors}
\author{J. Linder}
\affiliation{Department of Physics, Norwegian University of
Science and Technology, N-7491 Trondheim, Norway.}
\author{A. Sudb{\o}}
\affiliation{Department of Physics, Norwegian University of
Science and Technology, N-7491 Trondheim, Norway.}
\affiliation{Centre for Advanced Study, Norwegian Academy of Science and Letters, Drammensveien 78, N-0271 Oslo, Norway.  }
\date{Received \today}
\begin{abstract}
Motivated by recent findings of unconventional superconductors exhibiting multiple broken symmetries, we consider a general Hamiltonian describing coexistence of itinerant ferromagnetism,  spin-orbit coupling and mixed spin-singlet/triplet superconducting pairing in the context of mean-field theory. The Hamiltonian is diagonalized and exact eigenvalues are obtained, thus allowing us to write down the coupled gap equations for the different order parameters. Our results may then be applied to any model describing coexistence of any combination of these three phenomena. As a specific application of our results, we consider tunneling between a normal metal and a noncentrosymmetric superconductor with mixed singlet and triplet gaps. The conductance spectrum reveals information about these gaps in addition to how the influence of spin-orbit coupling is manifested. Explicitly, we find well-pronounced peaks and bumps in the spectrum at voltages corresponding to the sum and the difference of the magnitude of the singlet and triplet components. Our results may thus be helpful in determining the relative sizes of the singlet and triplet gaps in noncentrosymmetric superconductors. We also consider the coexistence of itinerant ferromagnetism and triplet superconductivity as a model for recently discovered ferromagnetic superconductors. The coupled gap equations are solved self-consistently, and we study the conditions necessary to obtain the coexistent regime of ferromagnetism and superconductivity. Analytical expressions are presented for the order parameters, and we provide an analysis of the free energy to identify the preferred system state. It is found that the uniform coexistence of ferromagnetism and superconductivity is energetically favored compared to both the purely ferromagnetic state and the unitary superconducting state with zero magnetization. Moreover, we make specific predictions concerning the heat capacity for a ferromagnetic superconductor. In particular, we report a nonuniversal relative jump
in the specific heat, depending on the magnetization of the system, at the uppermost superconducting phase transition. We propose that this may be exploited to obtain information about both the superconducting pairing symmetry realized in ferromagnetic superconductors in addition to the magnitude of the exchange splitting between majority and minority spin bands.  
  \end{abstract}
\pacs{74.20.-z, 74.25.-q, 74.45.+c, 74.50.+r, 74.20.Rp}

\maketitle

\section{Introduction}
Recent findings of superconductors that 
simultaneously exhibit multiple spontaneously broken 
symmetries, such as ferromagnetic order or lack of an 
inversion center \cite{saxena,aoki,bauer1} and even 
combinations of such broken symmetries \cite{akazawa1}, 
have led to much theoretical and experimental research \cite{huxley,samokhin,machida}. 
The symmetry of the superconducting 
gap in these and other unconventional superconductors is 
presently a matter of intense 
investigation \cite{nelson,yuan,curro,lebed,mazin}. Multiple spontaneously broken symmetries are not only of interest in terms of studying properties of specific condensed matter systems, but also due to the fact that it may provide clues 
for what could be expected in other systems in vastly 
different areas of physics. Topics such as mass-differences of elementary particles and emergent phenomena in biology is caused by spontaneously broken symmetries \cite{andersonbook}, and in many cases, the phenomena may even be described by the same type of equations. In this paper, we will address the issue of competition and coexistence between three phenomena giving rise to broken symmetries which are highly relevant in condensed-matter physics: ferromagnetism, superconductivity, and spin-orbit coupling. 
\\
\indent The discovery of superconducting materials that lack a centre of inversion \cite{bauer1, yogi, akazawa1, sergienko, yuan}, such as CePt$_3$Si, UIr, Li$_2$Pd$_3$B, Li$_2$Pt$_3$B, and Cd$_2$Re$_2$O$_7$, has lately triggered  extensive theoretical work on these compunds. Properties of a superconductor without an inversion center were investigated early by Edelstein \cite{edelstein}, while in Ref. \onlinecite{gorkov} it was  shown that a 2D superconducting system with a significant spin-orbit coupling induced by the lack of inversion symmetry would display a mixed singlet-triplet superconducting state. This means that the superconducting order parameter would possess the exotic feature of having no definite parity. Later studies \cite{sergienko2, borkje, frigeri2} also investigated specific noncentrosymmetric superconductors with a model Hamiltonian consisting of a superposition of spin-orbit and superconducting terms. In an attempt to determine the correct pairing symmetry of the superconducting state in such unconventional superconductors, it was found that the favored triplet pairing state \cite{frigeri} for the heavy-fermion material CePt$_3$Si is $\mathbf{d}_\vk \propto (k_y,-k_x,0)$. Very recently, however, an experimental study \cite{izawa} of thermal transport properties in the present compound concluded that the correct gap function ($\mathbf{d}_\vk$-vector) may exhibit nodal lines in contrast to the point nodes displayed by the $\mathbf{d}_\vk$-vector suggested by Ref.~\onlinecite{frigeri}. It is therefore of considerable interest to investigate several specific models for noncentrosymmetric superconductors in order to reveal characteristic features in physical observables that might be helpful in classifying the symmetry of the superconducting order parameter. \\
\indent In Ref.~\onlinecite{tanaka2}, the authors studied tunneling between a normal metal and a noncentrosymmetric superconductor considering the particular form of $\mathbf{d}_\vk$ suggested by Ref.~\onlinecite{frigeri} in the limit of weak spin-orbit coupling and in the absence of spin-singlet pairing. Anderson \cite{anderson} showed that the only stable triplet pairing states in the presence of a spin-orbit coupling would have to satisfy $\mathbf{d}_\vk \parallel \mathbf{g}_\vk$, where $\mathbf{g}_\vk = -\mathbf{g}_{-\vk}$ is the vector function describing this interaction, such that in CePt$_3$Si one also has $\mathbf{g}_\vk = \lambda(k_y,-k_x,0)$. Moreover, it was demonstrated by Samokhin \cite{samokhin2} that the spin-orbit coupling in this particular material is significant, \ie $\gg k_BT_c$, which indicates admixturing of singlet and triplet Cooper pairs. In the present paper, we solve the full Bogoliubov-de Gennes (BdG)-equations for 
a system with spin-orbit coupling including both spin-singlet and spin-triplet superconducting gaps, studying a gap vector $\mathbf{d}_{\vk,\text{point}}\propto (k_y,-k_x,0)$ as suggested by Ref.~\onlinecite{frigeri}. We then apply this gap vector to what we believe is the simplest model that captures the essential features that could be expected to appear in the conductance spectrum of a 2D normal/CePt$_3$Si junction. Our work then significantly extends the considerations made in Ref.~\onlinecite{tanaka2} primarily in that we present analytical and numerical results that allow for \textit{both} triplet and singlet gap components. Also note that a similar Hamiltonian was very recently studied in Ref.~\onlinecite{eremin2}, where it was shown that the presence of a weak external magnetic field would significantly change the nodal topology of CePt$_3$Si. With regard to noncentrosymmetricity, we underline that breaking the symmetry of spatial inversion does not in general give rise to a significant spin-orbit coupling. Also, it is well-known that spin-orbit coupling may be induced in a centrosymmetric crystal by means of an external symmetry-breaking electrical field. In the latter case, however, the broken symmetry is strictly speaking not spontaneous as it certainly is for \eg a crystal lattice undergoing a structural phase transition which breaks spatial inversion \cite{sergienko}. \\
\indent Another interesting scenario in the context of spontaneously broken symmetries is the study of superconductors that exhibit coexistence of ferromagnetic
and superconducting order, i.e. systems where two continuous internal symmetries $SU(2)$ and $U(1)$ are simultaneously broken. Due to the preferred orientation of the spins in a ferromagnetic system, the $SU(2)$ rotational symmetry is spontaneously broken. In a superconducting system, the ground state spontaneously breaks the $U(1)$ symmetry. Note that by the terminology broken symmetry, we are referring to the fact that the wavefunction describing the state of the system acquires a complex phase which characterizes the ground state.
In the ferromagnetic and superconducting systems we will consider in this paper,
superconductivity appears at a lower temperature than the temperature at which onset of 
ferromagnetism is found. This may be simply due to the fact that the energy scales for the two phenomena 
are quite different, with the exchange energy naturally being the largest. It may, however, also be due to 
the fact that superconductivity is dependent on ferromagnetism for its very existence. 
Such a suggestion has recently been put forth \cite{niu}. \\
\indent In the context of FMSCs, it is crucial to address the question of whether the superconductivity and ferromagnetism
order parameters coexist uniformly or if they are phase-separated. One plausible scenario 
\cite{tewari2004} is that a spontaneously formed vortex
lattice due to the internal magnetization $\mathbf{M}$ is realized, but studies of a uniform superconducting phase in spin-triplet FMSCs \cite{shopova2005} has also been conducted. As argued
by Mineev in Ref.~\onlinecite{mineev2005}, an important factor with respect to
whether a vortice lattice appears or not should be the magnitude of the
internal magnetization $\mathbf{M}$. Specifically, Ref.~\onlinecite{mineev1999}
suggested that vortices may arise if $4\pi\mathbf{M}>\mathbf{H}_{c1}$, where
$\mathbf{H}_{c1}$ is the lower critical field. In the case of
URhGe, a weakly
ferromagnetic state coexisting with superconductivity seems to be realized, and the domain structure in the absence of an external
field is thus vortex-free. Unfortunately, current experimental data concerning
URhGe are not as of yet strong enough to unambiguously settle this
question. On the other hand, evidence for uniform coexistence of ferromagnetism and superconductivity has been
indicated \cite{kotegawa2005} in UGe$_2$. \\
\indent Although this is an unsettled issue, it seems natural to assume that in ferromagnetic 
superconductors (FMSCs), the electrons involved in the $SU(2)$ symmetry breaking also participate 
in the $U(1)$ symmetry breaking.  As a consequence, uniform coexistence of spin-singlet superconductivity 
and ferromagnetism can be discarded since $s$-wave Cooper pairs carry a total spin of zero, although 
spatially modulated order parameters could allow for magnetic $s$-wave
superconductors \cite{kulic2005,eremin2006}. However, spin-triplet
Cooper pairs are in principle perfectly compatible with ferromagnetic order since
they can carry a net magnetic moment. There is strong reason to
believe that the correct pairing symmetries in the discovered FMSCs
constitute non-unitary states \cite{hardy2005,samokhin2002}. Spin-triplet superconductors have a multicomponent order parameter $\mathbf{d}_\vk$, which for a given spin basis reads
\begin{equation}\label{eq:dvector}
\mathbf{d}_\vk = \Big[\frac{\Delta_{\vk\downarrow\downarrow} - \Delta_{\vk\uparrow\uparrow}}{2}, \frac{-\i(\Delta_{\vk\downarrow}+\Delta_{\vk\uparrow\uparrow})}{2}, \Delta_{\vk\uparrow\downarrow}\Big].
\end{equation}
Note that $\mathbf{d}_\vk$ transforms like a vector under spin rotations. The superconducting order parameter is characterized as non-unitary if $\i(\mathbf{d}_\vk\times\mathbf{d}_\vk^*) \neq 0$, which effectively means that time-reversal symmetry is broken in the spin part of the Cooper pairs, since the average spin of Cooper pairs is given as
$\langle \mathbf{S}_\vk \rangle = \i(\mathbf{d}_\vk\times\mathbf{d}_\vk^*)$. Notice that time-reversal symmetry may be broken in the orbital part (angular momentum) of the Cooper pair wavefunction even if the state is unitary. In the general case where all SC gaps are included, it is generally argued that $\Delta_{\vk\uparrow\downarrow}$ would be suppressed in the presence of a Zeeman-splitting between the $\uparrow,\downarrow$ conduction bands. Distinguishing between unitary and non-unitary states in FMSCs is clearly one of the primary objectives in terms of identifying the correct order parameter. Studies of quantum transport in junctions involving FMSCs has explicitly shown that the conductance spectrum should be helpful in revealing the correct pairing symmetry \cite{linder, yokoyama07}.
Hence, an itinerant electron model of ferromagnetism augmented by a suitable pairing kernel should be 
a reasonable starting point for describing such systems. 
\par
Although we have mentioned two specific examples of systems exhibiting multiple broken symmetries, our aim with this paper is to construct a solid starting point for consideration of a condensed-matter system exhibiting any combination of the broken symmetries resulting from superconductivity, ferromagnetism, and/or spin-orbit coupling. By applying the appropriate limits to our theory, one may then obtain special cases such as FMSCs or noncentrosymmetric superconductors with significant spin-orbit coupling.\\
\indent This paper is organized as follows. In Sec. \ref{sec:model}, we establish the Hamiltonian accounting 
for general coexistence of ferromagnetism, spin-orbit coupling, and superconductivity. The diagonalization 
procedure and coupled gap equations are described in Sec. \ref{sec:eigen}. Then, we apply our findings to 
a model of normal/noncentrosymmetric superconductor junction, calculating the tunneling conductance 
spectrum Sec. \ref{sec:noncen}, in addition to a discussion of these results. As a second application, we 
consider a FMSC in Sec. \ref{sec:fermag}, solving the coupled gap equations 
self-consistently and calculating the free energy and heat capacity of such a system. Our main conclusions are summarized in Sec. \ref{sec:summary}. We will use boldface notation for vectors, $\hat{...}$ for operators, and $\check{...}$ 
for 2$\times$2 matrices.

\section{Model for coexistence of ferromagnetism, spin-orbit coupling, and superconductivity}\label{sec:model}
For our model, we will write down a Hamiltonian  describing 
the kinetic energy, exchange energy, spin-orbit coupling, and attractive electron-electron interaction, 
respectively. The total Hamiltonian can then be written as
\begin{equation}
\hat{H} = \hat{H}_\text{kin} + \hat{H}_\text{FM} + \hat{H}_\text{SOC} + \hat{H}_\text{SC},
\end{equation}
where the respective individual terms read
\begin{align}\label{eq:kspace}
\hat{H}_\text{kin} &= \sum_{\vk\sigma} \xi_\vk \cop_{\vk\sigma}^\dag \cop_{\vk\sigma},\notag\\
\hat{H}_\text{FM} &= -JN\sum_\vk\gamma(\vk) \hat{\mathbf{S}}_\vk \cdot\hat{\mathbf{S}}_{-\vk},\notag\\
\hat{H}_\text{SOC} &= \sum_{\vk\alpha\beta} \cop_{\vk\alpha}^\dag (\vg_\vk\cdot \check{\boldsymbol{\sigma}})_{\alpha\beta}\cop_{\vk\beta}, \notag\\
\hat{H}_\text{SC} &= \frac{1}{2N}\sum_{\vk\vk'\alpha\beta}(V_{\vk\vk'\alpha\beta}^\text{S} + V_{\vk\vk'\alpha\beta}^\text{T})\cop_{\vk\alpha}^\dag \cop_{-\vk\beta}^\dag \cop_{-\vk'\beta} \cop_{\vk'\alpha}.
\end{align}
Above, $\xi_\vk = \varepsilon_\vk - \mu$ where $\varepsilon_\vk$ is the dispersion relation for the free fermions and $\mu$ is the chemical potential \footnote{At zero temperature, the chemical potential is identically equal to the Fermi energy. In this context, we introduce it as a reference point for energy measurements such that the single-particle kinetic energies are measured relative $\mu$.}, $J>0$ is a ferromagnetic coupling parameter, $\gamma(\vk)$ is a geometrical structure factor for the lattice, $\mathbf{g}_\vk$ is a vector function accounting for the antisymmetric spin-orbit coupling, while $V_{\vk\vk'\alpha\beta}$ is an attractive pair potential. The factor of $1/2$ in $\hat{H}_\text{SC}$ is included to obtain more convenient expressions later on, and simply corresponds to a redefinition of $V_{\vk\vk'\alpha\beta}^\text{S,T} \to \frac{1}{2} V_{\vk\vk'\alpha\beta}^\text{S,T}$. In Eqs. (\ref{eq:kspace}), the spin operators are given by
\begin{align}
\hat{\mathbf{S}}_\vk &= \frac{1}{N} \sum_{\vk'} \cop_{\vk'\alpha}^\dag \check{\boldsymbol{\sigma}}_{\alpha\beta} \cop_{(\vk+\vk')\beta}.
\end{align}
Moreover, we have explicitly split the attractive pairing potential into a singlet and triplet part 
according to $V_{\vk\vk'\alpha\beta} = V_{\vk\vk'\alpha\beta}^\text{S} + V_{\vk\vk'\alpha\beta}^\text{T}$. 
The symmetry properties of the antisymmetric spin-orbit coupling and superconductivity terms with respect 
to spatial inversion symmetry read
\begin{align}
\vg_{\vk} &= -\vg_{-\vk},\; V^\text{S}_{\vk\vk'\alpha\beta} = V^\text{S}_{\pm\vk\pm'\vk'\alpha\beta},\notag\\ V^\text{T}_{\vk\vk'\alpha\beta} &= \pm\pm'V^\text{T}_{\pm\vk\pm'\vk'\alpha\beta}.
\end{align}
In order to find eigenvalues and gap equations for our system, we introduce the mean-field approximation for the two-particle Hamiltonians (ferromagnetic and superconducting terms) such that the operators $\hat{\mathbf{S}}_\vk$ and $\cop_{\vk\alpha}^\dag \cop_{-\vk\beta}^\dag$ may be written as a mean-field value pluss small fluctuations. We define $\langle \cop_{\vk\alpha}^\dag \cop_{-\vk\beta}^\dag \rangle = b^\dag_{\vk\alpha\beta}$, and write
\begin{align}\label{eq:meanfield}
\hat{\mathbf{S}}_\vk &= \langle \hat{\mathbf{S}}_\vk \rangle + \delta\langle\hat{\mathbf{S}}_\vk\rangle,\notag\\
\cop_{\vk\alpha}^\dag \cop_{-\vk\beta}^\dag &= b^\dag_{\vk\alpha\beta} +\delta b^\dag_{\vk\alpha\beta}. 
\end{align}
Inserting Eqs. (\ref{eq:meanfield}) into Eqs. (\ref{eq:kspace}) and discarding all terms of order ${\cal{O}}(\delta^2)$, one obtains in the standard fashion
\begin{align}\label{eq:meanfieldeqs}
\hat{H}_\text{FM} &= -\sum_{\vk\alpha\beta} \cop_{\vk\alpha}^\dag (\vV_M\cdot\check{\boldsymbol{\sigma}})_{\alpha\beta}\cop_{\vk\beta} + \frac{INM^2}{2},\notag\\
\hat{H}_\text{SC} &= \frac{1}{2}\sum_{\vk\alpha\beta}\Big\{
[(\Delta_{\vk\alpha\beta}^\text{S})^\dag + (\Delta_{\vk\alpha\beta}^\text{T})^\dag]\cop_{-\vk\beta}\cop_{\vk\alpha} \notag\\
& +  [\Delta_{\vk\alpha\beta}^\text{S} + \Delta_{\vk\alpha\beta}^\text{T}][\cop_{\vk\alpha}^\dag \cop_{-\vk\beta}^\dag - b_{\vk\alpha\beta}^\dag\Big\}.
\end{align}
In Eqs. (\ref{eq:meanfieldeqs}), $\mathbf{M} = (M_x,M_y,M_z) = \langle \hat{\mathbf{S}}_i \rangle = \langle \hat{\mathbf{S}}_{\vk=0} \rangle$ denotes the mean value of the spin operators in real space, interpreted as the magnetization of the system. We have introduced the vector describing the magnetic exchange energy $\vV_M = I\mathbf{M}$ and the order parameters (OPs) 
\begin{align}\label{eq:zeta}
\mathcal{V} = (\vV_M)_x - \i (\vV_M)_y = I(M_x - \i M_y),\; \mathcal{V}_z = IM_z,
\end{align}
for ferromagnetism, while the OP for superconductivity is described by
\begin{align}
\Delta_{\vk\alpha\beta}^\text{S,T} &= \frac{1}{N}\sum_{\vk'} V_{\vk\vk'\alpha\beta}^\text{S,T} b_{\vk'\alpha\beta},\notag\\
\Big(\Delta_{\vk'\alpha\beta}^\text{S,T}\Big)^\dag &= \frac{1}{N}\sum_{\vk} V_{\vk\vk'\alpha\beta}^\text{S,T} b_{\vk\alpha\beta}^\dag.
\end{align}
The quantity $I$ appearing in Eq. (\ref{eq:zeta}) is a measure of the strength of the magnetic exchange coupling. Although we have derived the ferromagnetic part of our Hamiltonian from a lattice-model [where $I = 2JN\gamma(0)$], this generic Hamiltonian describes a general mean-field model of a system with magnetic exchange energy. The Pauli principle places the following restrictions upon the superconductivity OPs:
\begin{align}
\text{Singlet pairing: }\;  &\Delta_{\vk\alpha\beta}^\text{S} = -\Delta_{\vk\beta\alpha}^\text{S},\; \Delta_{\vk\alpha\beta}^\text{S} = \Delta_{-\vk\alpha\beta}^\text{S}. \notag\\
\text{Triplet pairing: }\;  &\Delta_{\vk\alpha\beta}^\text{T} = \Delta_{\vk\beta\alpha}^\text{T},\; \Delta_{\vk\alpha\beta}^\text{T} = -\Delta_{-\vk\alpha\beta}^\text{T}. \notag\\
\end{align}
In total, we have thus obtained a Hamiltonian $\hat{H}$ describing coexistence of ferromagnetism, spin-orbit 
coupling, and superconductivity in the mean-field approximation by adding all of the above terms. For more 
compact notation, one may introduce a basis for fermion operators $\hat{\phi}_\vk = [\cop_{\vk\uparrow}, \cop_{\vk\downarrow}, \cop_{-\vk\uparrow}^\dag, \cop_{-\vk\downarrow}^\dag]^\text{T}$ and write
\begin{align}\label{eq:Hold}
\hat{H} &= \hat{H}_\text{kin} + \hat{H}_\text{FM} + \hat{H}_\text{SOC} + \hat{H}_\text{SC}  \notag\\
&= H_0 + \frac{1}{2}\sum_{\vk}\hat{\phi}_\vk^\dag \A\hat{\phi}_\vk,
\end{align}
where we have introduced the quantities
\begin{widetext}
\begin{align}\label{eq:Amatrix}
H_0 &= \sum_{\vk}\xi_\vk + \frac{INM^2}{2} - \frac{N(\mathcal{V}+\mathcal{V}^\dag)}{2}  - \frac{1}{2} \sum_{\vk\alpha\beta} (\Delta_{\vk\alpha\beta}^\text{S} + \Delta_{\vk\alpha\beta}^\text{T})b_{\vk\alpha\beta}^\dag,\notag\\
\A &= \begin{pmatrix}
\xi_{\vk\uparrow} + g_{\vk,z} & -\mathcal{V} + g_{\vk,-} &  \Delta_{\vk\uparrow\uparrow}^\text{T} & \Delta_{\vk\uparrow\downarrow}^\text{S} + \Delta_{\vk\uparrow\downarrow}^\text{T}\\
-\mathcal{V}^\dag + g_{\vk,+} & \xi_{\vk\downarrow} - g_{\vk,z} & -\Delta_{\vk\uparrow\downarrow}^\text{S} + \Delta_{\vk\uparrow\downarrow}^\text{T} & \Delta_{\vk\downarrow\downarrow}^\text{T}\\
(\Delta_{\vk\uparrow\uparrow}^\text{T})^\dag & (\Delta_{\vk\uparrow\downarrow}^\text{T})^\dag - (\Delta_{\vk\uparrow\downarrow}^\text{S})^\dag & -\xi_{\vk\uparrow} + g_{\vk,z} & \mathcal{V}^\dag + g_{\vk,+} \\
(\Delta_{\vk\uparrow\downarrow}^\text{T})^\dag + (\Delta_{\vk\uparrow\downarrow}^\text{S})^\dag & (\Delta_{\vk\downarrow\downarrow}^\text{T})^\dag & \mathcal{V} + g_{\vk,-} & -\xi_{\vk\downarrow} - g_{\vk,z} \\
\end{pmatrix}.
\end{align}
\end{widetext}
Above, we have defined $\xi_{\vk\sigma} = \xi_\vk - \sigma\mathcal{V}_z$ in addition to $g_{\vk,\pm} = (\mathbf{g}_\vk)_x \pm \i(\mathbf{g}_\vk)_y$. The matrix $\A$ will be central in this work, and we note that 
it may be further compactified by introducing the $\mathbf{d}_\vk$-vector formalism \cite{leggett1975}. By 
means of the definitions $d_{\vk,0} = \Delta_{\vk\uparrow\downarrow}^\text{S}$ and
\begin{equation}
\mathbf{d}_\vk = \frac{1}{2}[\Delta_{\vk\downarrow\downarrow}^\text{T} - \Delta_{\vk\uparrow\uparrow}^\text{T}, -\i(\Delta_{\vk\uparrow\uparrow}^\text{T} + \Delta_{\vk\downarrow\downarrow}^\text{T}), 2\Delta_{\vk\uparrow\downarrow}^\text{T}],
\end{equation}
that transforms like a vector under spin rotations, one may write
\begin{equation}
\A = \begin{pmatrix}
\xi_\vk \check{\mathbf{1}} - \check{\boldsymbol{\sigma}}\cdot(\vV_M - \mathbf{g}_\vk) & \i(d_{0,\vk} + \mathbf{d}_\vk\cdot\check{\boldsymbol{\sigma}})\check{\sigma}_y \\
[\i(d_{0,\vk} + \mathbf{d}_\vk\cdot\check{\boldsymbol{\sigma}})\check{\sigma}_y]^\dag & -\xi_\vk \check{\mathbf{1}} + (\vV_M + \mathbf{g}_\vk)\cdot\check{\boldsymbol{\sigma}}^{\cal{T}}  \\
\end{pmatrix},
\end{equation}
where $\check{\mathbf{1}}$ denotes the identity matrix and ${\cal{T}}$ designates the matrix transpose.
The rest of this paper will now be devoted to obtaining the excitation energies for $\hat{H}$ by 
diagonalizing $\A$, writing down the coupled gap equations, and considering some important special 
cases.

\section{Excitation energies and gap equations}\label{sec:eigen}
The characteristic polynomial for a general matrix $A_\vk$ with eigenvalues $E_\vk$ may be written as \cite{edwards} 
\begin{align}\label{eq:poly1}
\rho(E_\vk) &= E_\vk^4 - (\text{Tr}\{A_\vk\})E_\vk^3 \notag\\
&+ \{\frac{1}{2}[|A_\vk-I| + |A_\vk+I|] - 1 - \text{det}A_\vk\}E_\vk^2 \notag\\
&+ \frac{1}{2}[|A_\vk-I| - |A_\vk+I|]E_\vk +  \text{det}A_\vk = 0,
\end{align}
where $I$ denotes the 4$\times$4 identity matrix. Since $A_\vk$ in our case is Hermitian, Tr$\{A_\vk\}=0$, and the polynomial reduces to a depressed quartic equation. For ease of notation, we introduce the quantity
\begin{equation}
\eta_\vk^\pm \equiv \frac{1}{2}[|A_\vk-I| \pm |A_\vk+I|],
\end{equation}
such that Eq. (\ref{eq:poly1}) is rewritten as
\begin{equation}\label{eq:poly2}
E_\vk^4 + (\eta_\vk^+ - 1 - \text{det}A_\vk)E_\vk^2 + \eta_\vk^-E_\vk + \text{det}A_\vk =0.
\end{equation}
The solutions of $E_\vk$ can be written as \cite{abramowitz}
\begin{align}\label{eq:tausol}
2E_{\vk_{\alpha\beta}} &= \alpha a_\vk +\beta\{-[3(\eta_\vk^+ - 1 - \text{det}A_\vk) \notag\\
&+ 2y_\vk +\alpha \frac{2\eta_\vk^-}{a_\vk}]\}^{1/2}.
\end{align}
Here, we have defined the auxiliary quantities
\begin{align}
a_\vk &= \sqrt{(\eta_\vk^+ - 1 - \text{det}A_\vk) + 2y_\vk},\notag\\
y_\vk &= -\frac{5(\eta_\vk^+ - 1 - \text{det}A_\vk)}{6} - b_\vk,\notag\\
b_\vk &= R_\vk^{1/3},\; R_\vk = \frac{Q_\vk}{2} + \sqrt{\frac{Q_\vk^2}{4} + \frac{P_\vk^3}{27}},\notag\\
Q_\vk &= \frac{(\eta_\vk^+ - 1 - \text{det}A_\vk)\text{det}A_\vk}{3} - \frac{(\eta_\vk^-)^2}{8}\notag\\
&-\frac{(\eta_\vk^+ - 1 - \text{det}A_\vk)^3}{108},\notag\\
P_\vk &= -\frac{(\eta_\vk^+ - 1 - \text{det}A_\vk)^2}{12} - \text{det}A_\vk.
\end{align}
In Eq. (\ref{eq:tausol}), $\{\alpha,\beta\}$ take the values $+1$ and $-1$ such that there exists a total of four solutions for $E_\vk$. Also note that any of the roots in the expressions for $b_\vk$ and $R_\vk$ will do the job. A special case of the above solutions, which occurs quite frequently in various contexts, considerably simplifies the obtained eigenvalues: $\frac{1}{2}[|A_\vk-I| - |A_\vk+I|] = 0$. In this case, the quartic equation reduces to an effective quadratic equation with the solutions
\begin{align}
2E_{\vk_{\alpha\beta}} &= \alpha\Big[-2(\eta_\vk^+-1-\text{det}A_\vk) \notag\\
&+ 2\beta\sqrt{(\eta_\vk^+-1-\text{det}A_\vk)^2 - 4\text{det}A_\vk}\Big]^{1/2}.
\end{align}
This is the situation considered in most problems dealing with superconductors.
Having calculated the energy eigenvalues, Eq. (\ref{eq:Hold}) may now be diagonalized by writing
\begin{align}\label{eq:Hny}
\hat{H} &= H_0 + \frac{1}{2}\sum_\vk \hat{\phi}_\vk^\dag \A \hat{\phi}_\vk\notag\\
&= H_0 + \frac{1}{2}\sum_\vk (\hat{\phi}_\vk^\dag P_\vk) (P_\vk^\dag \A P_\vk) (P_\vk^\dag\hat{\phi}_\vk)\notag\\
&= H_0 + \sum_\vk \hat{\widetilde{\phi}}_\vk^\dag D_\vk \hat{\widetilde{\phi}}_\vk,
\end{align}
where $D_\vk = \text{diag}(E_{\vk,1}, E_{\vk,2}, E_{\vk,3}, E_{\vk,4})$ is a diagonal matrix containing the eigenvalues of $\A$. Here, we have defined [see Eq. (\ref{eq:tausol})]
\begin{align}
E_{\vk,1} &= \frac{1}{2}E_{\vk_{++}},\; \frac{1}{2}E_{\vk,2} = E_{\vk_{+-}}, \notag\\
E_{\vk,3} &= \frac{1}{2}E_{\vk_{-+}},\; \frac{1}{2}E_{\vk,4} = E_{\vk_{--}},
\end{align}
thus absorbing the factor $\frac{1}{2}$ in front of $\sum_\vk$ into the eigenvalues.
Above, $P_\vk$ are the orthonormal diagonalizing matrices which by the hermiticity of $\A$ are ensured to be unitary. We write our new basis of fermion operators as 
\begin{equation}
\hat{\widetilde{\phi}}_\vk^\dag = [\hat{\gamma}_{\vk\uparrow}^\dag, \hat{\gamma}_{\vk\downarrow}^\dag, \hat{\gamma}_{-\vk\uparrow}, \hat{\gamma}_{-\vk\downarrow}].
\end{equation}
 These operators satisfy the fermion anticommutation relations, as can be verified by direct insertion. From Eq. (\ref{eq:Hny}), we may now write
\begin{align}
\hat{H} &= H_0 + \sum_\vk[\gamma_{\vk\uparrow}^\dag\gamma_{\vk\uparrow}E_{\vk,1} + \gamma_{\vk\downarrow}^\dag\gamma_{\vk\downarrow}E_{\vk,2}  \notag\\
&\hspace{0.6in}+ \gamma_{-\vk\uparrow}\gamma_{-\vk\uparrow}^\dag E_{\vk,3} + \gamma_{-\vk\downarrow}\gamma_{-\vk\downarrow}^\dag E_{\vk,4}]\notag\\
&= [H_0 +\sum_{\vk}(E_{\vk,3} + E_{\vk,4})] \notag\\
&\hspace{0.4in}+ \sum_\vk[\gamma_{\vk\uparrow}^\dag\gamma_{\vk\uparrow}(E_{\vk,1} - E_{-\vk,3}) \notag\\
&\hspace{0.6in}+ \gamma_{\vk\downarrow}^\dag\gamma_{\vk\downarrow}(E_{\vk,2} -E_{-\vk,4})]\notag\\
&= \widetilde{H}_0 + \sum_\vk[\gamma_{\vk\uparrow}^\dag\gamma_{\vk\uparrow}\widetilde{E}_{\vk,1} + \gamma_{\vk\downarrow}^\dag\gamma_{\vk\downarrow}\widetilde{E}_{\vk,2}],
\end{align}
where we have defined $\widetilde{H}_0 = H_0 + \sum_{\vk}(E_{\vk,3} + E_{\vk,4})$ and $\widetilde{E}_{\vk,1} = (E_{\vk,1} - E_{-\vk,3})$, $\widetilde{E}_{\vk,2} = (E_{\vk,2} - E_{-\vk,4})$. Our Hamiltonian now has the form of a free-fermion theory. 
It is then readily seen that the free energy of the system is given by
\begin{equation}\label{eq:F}
F = \widetilde{H}_0 - \frac{1}{\beta}\sum_\vk[ \text{ln}(1+\e{-\beta\widetilde{E}_{\vk,1}}) + \text{ln}(1+\e{-\beta\widetilde{E}_{\vk,2}})].
\end{equation}
From $F$, the gap equations for the ferromagnetic and superconducting OPs $\mathcal{V}, \mathcal{V}_z$, and $\Delta_{\vk\alpha\beta}^\text{S,T}$ may be obtained by demanding the value of these which corresponds to a minimum in $F$. The possible extrema of $F$ are given by the conditions 
\begin{align}\label{eq:conditionsF}
\frac{\partial F}{\partial \Delta_{\vk\alpha\beta}^{S,T}} = 0,\; \frac{\partial F}{\partial (\Delta_{\vk\alpha\beta}^{S,T})^\dag} = 0,\notag\\
\frac{\partial F}{\partial \mathcal{V}_z} = 0,\; \frac{\partial F}{\partial \mathcal{V}} = 0,\;  \frac{\partial F}{\partial \mathcal{V}^\dag} = 0.
\end{align}
By first defining the quantity
\begin{align}\label{eq:F(x)}
{\cal{F}}(x) = \sum_\vk[&n_\text{F}(\widetilde{E}_{\vk,1})\frac{\partial \widetilde{E}_{\vk,1}}{\partial x} + \frac{\partial E_{\vk,3}}{\partial x}   \notag\\
&+ n_\text{F}(\widetilde{E}_{\vk,2})\frac{\partial \widetilde{E}_{\vk,2}}{\partial x}    + \frac{\partial E_{\vk,4}}{\partial x}],
\end{align}
where $n_\text{F}(E) = 1/(1+\e{\beta E})$ is the Fermi distribution, the conditions in Eqs. (\ref{eq:conditionsF}) may be evaluated by inserting Eq. (\ref{eq:F}). The extrema of $F$ are thus determined by the following equations:
\begin{align}\label{eq:Fcond}
&-b_{\vk\alpha\beta}^\dag + {\cal{F}}(\Delta_{\vk\alpha\beta}^\text{S,T}) = 0,\\
&-b_{\vk\alpha\beta} + {\cal{F}}[(\Delta_{\vk\alpha\beta}^\text{S,T})^\dag] = 0,\\
&\frac{N\mathcal{V}_z}{2J\gamma(0)} + {\cal{F}}(\mathcal{V}_z) = 0,\\
&-\frac{N}{2} + \frac{N\mathcal{V}^\dag}{2J\gamma(0)} + {\cal{F}}(\mathcal{V}) = 0,\\
&-\frac{N}{2} + \frac{N\mathcal{V}}{2J\gamma(0)} + {\cal{F}}(\mathcal{V}^\dag) = 0.
\end{align}
The challenge then lies in obtaining the derivatives of the energies $E_{\vk,i}$ with respect to the different order parameters. In the general case described by Eq. (\ref{eq:Amatrix}), this is a formidable task. Nevertheless, 
the above above provides a general framework which may serve as a starting point for any model considering the coexistence of ferromagnetism, spin-orbit coupling, and superconductivity. We will apply our findings onto a specific case which currently is a topic attracting much attention: noncentrosymmetric superconductors with significant spin-orbit coupling.

\section{Probing the pairing symmetry of noncentrosymmetric superconductors}\label{sec:noncen}
As an application of our model, we consider tunneling between a normal metal and a noncentrosymmetric superconductor treated in the spin-generalized Blonder-Tinkham-Klapwijk (BTK) formalism \cite{btk,tanaka}.

\subsection{Model and formulation}
The Hamiltonian in the superconducting state using standard mean-field theory with a spin-orbit coupling may be written as 
\begin{equation}\label{eq:H}
\hat{H} = H_0 + \frac{1}{2} \sum_\vk \hat{\phi}_\vk^\dag M_\vk \hat{\phi}_\vk,
\end{equation}
using a spin basis $\hat{\phi}_\vk = [\cop_{\vk\uparrow}, \cop_{\vk\downarrow}, \cop_{-\vk\uparrow}^\dag, \cop_{-\vk\downarrow}^\dag]^\text{T}$, and with
\begin{equation}\label{eq:M}
M_\vk = \begin{pmatrix}
\varepsilon_\vk & g_{\vk,-} & \Delta_{\vk\uparrow\uparrow} &  \Delta_\vk \\
g_{\vk,+} & \varepsilon_\vk & -\Delta_\vk & \Delta_{\vk\downarrow\downarrow} \\
\Delta_{\vk\uparrow\uparrow}^\dag &  -\Delta^\dag_\vk & -\varepsilon_\vk & g_{\vk,+} \\
\Delta^\dag_\vk & \Delta_{\vk\downarrow\downarrow}^\dag & g_{\vk,-}& -\varepsilon_\vk
\end{pmatrix}.
\end{equation}
In Eq. (\ref{eq:M}), all quantities have been defined in the previous section. It is usually argued that interband pairing in a noncentrosymmetric superconductors can be neglected due to a spin-split Fermi surface in the presence of spin-orbit coupling. This is motivated by realizing that the splitting could be as large as \cite{samokhin2} 50-200 meV for the noncentrosymmetric superconductor CePt$_3$Si, thus far greater than the superconducting critical temperature $k_BT_c\simeq 0.06$ meV in that compound. Accordingly, one might be tempted to also exclude the spin-singlet gap in the presence of a strong spin-orbit coupling motivated on physical grounds by the suppression of interband-pairing due to the spin-split Fermi surfaces. However, it is necessary to investigate the presence, although possibly small in magnitude, of a spin-singlet component of the gap to examine whether the conductance spectrum changes significantly in any respect compared to the scenario with exclusively triplet pairing. Another motivation for including the singlet gap is that the authors of Ref.~\onlinecite{frigeri} demonstrated that for small spin-orbit coupling, $\mathbf{d}_\vk \parallel \mathbf{g}_\vk$ yields the highest $T_C$ for CePt$_3$Si. This would thus correspond to a scenario where the triplet gap $\Delta_{\vk\uparrow\downarrow}$ is suppressed due to the above condition, although intraband-pairing is not strictly forbidden as a result of weak spin-orbit coupling, thus allowing for singlet pairing. \\
\indent Consider now a gap vector exhibiting point nodes. Since $\mathbf{d}_\vk$ in general is given by Eq. (\ref{eq:dvector}), the vector characterizing spin-orbit coupling $\vg_\vk = \lambda(k_y,-k_x,0)$ suggested by Ref.~\onlinecite{frigeri} results in
\begin{align}\label{eq:twogaps}
\Delta_{\vk\uparrow\uparrow} &= -\frac{\Delta_\text{t}}{2|\vk|}(k_y+\i k_x),\; \Delta_{\vk\downarrow\downarrow} = \frac{\Delta_\text{t}}{2|\vk|}(k_y-\i k_x).
\end{align} 
Diagonalization of the Hamiltonian in Eq. (\ref{eq:H}) yields eigenvalues and eigenvectors which are necessary to calculate the normal- and Andreev-reflection coefficients in a N/CePt$_3$Si junction. 
Assuming the simplest form of a $s$-wave superconducting gap that obeys the symmetry requirements dictated by the Pauli-principle, namely an isotropic gap $\Delta_\vk = \Delta_\text{s}$, we find that the eigenvalues of $M_\vk$ read
\begin{align}\label{eq:eigenvalues} 
E_{\vk_{\alpha\beta}} &= \alpha\sqrt{(\varepsilon + \beta|\vg_\vk|)^2 + |\Delta_\text{s} + \beta\Delta_\text{t}/2|^2}.
\end{align}
This is in complete agreement with the result of Ref.~\onlinecite{eremin2}. We are here assuming that all gaps have the same phase associated with the broken $U(1)$ gauge symmetry. In Eq. (\ref{eq:eigenvalues}), $\alpha=+(-)$ refers to electronlike (holelike) excitations, while $\beta=+(-)$ denotes the spin-orbit helicity index. The wavevectors may then be written as 
\begin{align}
q_e^\uparrow = q_h^\uparrow = \sqrt{k_F^2 + m^2\lambda^2} - m\lambda,\notag\\
q_e^\downarrow = q_h^\downarrow = \sqrt{k_F^2 + m^2\lambda^2} + m\lambda,
\end{align} 
when making the approximation that the magnitude of the superconducting gaps is small compared to the Fermi energy $\mu$ and considering the low-energy transport regime. Here, $k_\text{F}$ is the Fermi wave-vector. 
\par
\begin{figure}[h!]
\centering
\resizebox{0.5\textwidth}{!}{
\includegraphics{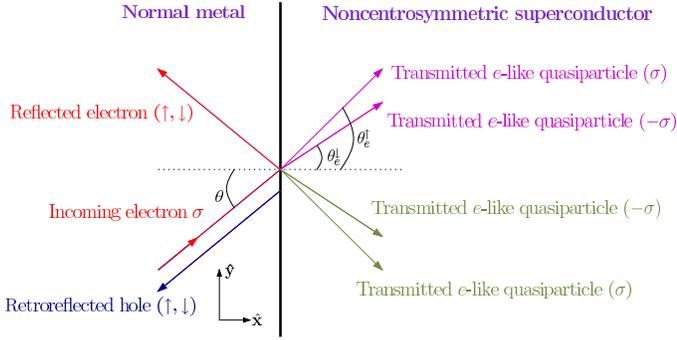}}
\caption{(Color online) Schematic illustration of the scattering processes taking place at the interface of the 2D planar N/CePt$_3$Si junction. The arrows indicate the direction of group velocity (which is not equal to the momentum vector for the holes). Note that the presence of spin-orbit coupling causes electron-like and hole-like excitations on the superconducting side to be spread into different angles.}
\label{fig:modelcentro}
\end{figure}

We now calculate the normal- and Andreev-reflection coefficients for an incident electron with spin $\sigma$, which in turn will allow us to derive the tunneling conductance of the junction. To do so, we first set up the 
Bogoliubov-de Gennes (BdG)-equations for the system which read (see Appendix A for a derivation):
\begin{widetext}
\begin{align}\label{eq:matrixnoncentro}
\begin{pmatrix}
-\frac{\hat{\nabla}^2}{2m} - \mu + V_0\delta(x) & \lambda(\hat{p}_y +\i\hat{p}_x)\Theta(x) & \Delta_{\vk\uparrow\uparrow}\Theta(x) & \Delta_\text{s}\Theta(x) \\
\lambda(\hat{p}_y-\i\hat{p}_x)\Theta(x) & -\frac{\hat{\nabla}^2}{2m} - \mu + V_0\delta(x)   & -\Delta_\text{s}\Theta(x) & \Delta_{\vk\downarrow\downarrow}\Theta(x) \\
\Delta_{\vk\uparrow\uparrow}^\dag\Theta(x) & -\Delta_\text{s}^\dag\Theta(x) & \frac{\hat{\nabla}^2}{2m} + \mu - V_0\delta(x) & \lambda(\hat{p}_y -\i\hat{p}_x)\Theta(x) \\
\Delta_\text{s}^\dag\Theta(x) & \Delta_{\vk\downarrow\downarrow}^\dag\Theta(x) & \lambda(\hat{p}_y +\i\hat{p}_x)\Theta(x) & \frac{\hat{\nabla}^2}{2m} + \mu - V_0\delta(x) \\
\end{pmatrix}
\Psi(x,y) = E\Psi(x,y),
\end{align}
\end{widetext}
where $\hat{p}_{x(y)} = -\i\hat{\partial}_{x(y)}$ and make use of the boundary conditions
\begin{align}\label{eq:boundary}
\textit{i)}\text{ }& \psi(0) = \Psi(0)\; \text{(Continuity of wavefunction)},\notag\\
\textit{ii)}\text{ }& 2mV_0\psi(0) = \hat{\partial}_x\Psi(0) - \hat{\partial}_x\psi'(0) - m\lambda\eta\Psi(0)\notag\\
& \text{(Continuity of flux)}.
\end{align}
Note that we have applied the usual step-function approximation for the order parameters instead of solving for their spatial dependence self-consistently near the interface, \ie $\Delta_{\vk\sigma\sigma}(\vecr) = \Delta_{\vk\sigma\sigma}\Theta(x)$ and $\Delta(\vecr) = \Delta\Theta(x)$ (we comment further on this later). For convenience, we have defined the 4$\times$4 matrix 
\begin{equation}
\eta = \begin{pmatrix} 
0 & 1 & 0 & 0 \\
-1 & 0 & 0 & 0 \\
0 & 0 & 0 & 1 \\
0 & 0 & -1 & 0 \\
\end{pmatrix}.
\end{equation}
The presence of spin-orbit coupling leads to off-diagonal components in the velocity operator, such that it would be erroneous to 
merely match the derivatives of the wavefunction in this case \cite{molenkamp}. 
The coupled gap equations that arise by demanding a minimum in the free energy are obtained by considering Eqs. (\ref{eq:F(x)}) and (\ref{eq:Fcond}). For the sake of obtaining analytical results, we continue our discussion of the conductance spectra of noncentrosymmetric superconductors by inserting values of the superconductivity gaps \textit{a priori} instead of using the self-consistent solutions. This approach does not, then, account for the entire physical picture, but has proven to yield satisfactory results for many aspects of quasiparticle tunneling in the case of \eg spin-singlet $d$-wave superconductors \cite{tanaka, tanaka97}.
\indent For the simplest model that illustrate the new physics, we have thus chosen a two-dimensional N/CePt$_3$Si junction with a barrier modelled by $V(\mathbf{r}) = V_0\delta(x)$ and superconductivity gaps $\Delta_{\vk\sigma\sigma}(\mathbf{r}) = \Delta_{\vk\sigma\sigma}\Theta(x)$ [$\delta(x)$ and $\Theta(x)$ represent the Delta- and Heaviside-function, respectively]. Consider Fig. \ref{fig:modelcentro} for an overview. Choosing a plane-wave solution $\Psi(x,y) = \Psi(x)\e{\i k_yy}$, for $\sigma=\uparrow$ the wavefunction on the normal $[\psi(x)]$ side of the junction reads
\begin{align}\label{eq:wavefunctions1}
\psi(x) &= 
\begin{pmatrix}
\e{\i k_\text{F}\cos\theta x} + r_e^\uparrow\e{-\i k_\text{F}\cos\theta x}\\
 r_e^\downarrow\e{-\i k_\text{F}\cos\theta x}\\
 r_h^\uparrow\e{\i k_\text{F}\cos\theta x}\\
  r_h^\downarrow\e{\i k_\text{F}\cos\theta x}\\
  \end{pmatrix}
\end{align}
On the superconducting side $[\Psi(x)]$, the BdG-equation may be written, for our particular choice of $\mathbf{g}_\vk$ and gaps in Eq. (\ref{eq:twogaps}), as
\begin{align}
\begin{pmatrix}
\varepsilon_\vk & |g_\vk|\e{\i\phi} & -(\Delta_\text{t}/2)\e{\i\phi} & \Delta_\text{s} \\
|g_\vk|\e{-\i\phi} & \varepsilon_\vk & -\Delta_\text{s} & (\Delta_\text{t}/2)\e{-\i\phi} \\
-(\Delta_\text{t}/2)\e{-\i\phi} & -\Delta_\text{s} & -\varepsilon_\vk & |g_\vk|\e{-\i\phi}\\
\Delta_\text{s} & (\Delta_\text{t}/2)\e{\i\phi} & |g_\vk|\e{\i\phi} & -\varepsilon_\vk\\
\end{pmatrix}\Psi \notag\\
= E\Psi,\; \text{with} \tan\phi(\theta) =  1/\tan\theta.
\end{align}
We are here concerned with positive excitations $E\geq0$, assuming an incident electron above Fermi level. In this case, there are four possible solutions for wavevectors $\vk$ with a given energy $E\geq 0$. Consequently, one may verify that the correct wavefunction for $x>0$, which is a linear combination of these allowed states, reads
\begin{widetext}
\begin{align}\label{eq:wavefunctions2}
\Psi(x) &= \frac{t_\text{e}^\uparrow}{\sqrt{2}}\begin{pmatrix}
u(\Delta_+)\\
u(\Delta_+)\e{-\i\phi(\theta_\text{e}^\uparrow)}\\
-v(\Delta_+)\e{-\i\phi(\theta_\text{e}^\uparrow)}\\
v(\Delta_+)\\
\end{pmatrix}\e{\i q_\text{e}^\uparrow\cos\theta_\text{e}^\uparrow x} 
+ 
\frac{t_\text{e}^\downarrow}{\sqrt{2}}\begin{pmatrix}
u(\Delta_-)\\
-u(\Delta_-)\e{-\i\phi(\theta_\text{e}^\downarrow)}\\
v(\Delta_-)\e{-\i\phi(\theta_\text{e}^\downarrow)}\\
v(\Delta_-)\\
\end{pmatrix}\e{\i q_\text{e}^\downarrow\cos\theta_\text{e}^\downarrow x}
\notag\\
&+
\frac{t_\text{h}^\uparrow}{\sqrt{2}}\begin{pmatrix}
v(\Delta_+)\\
v(\Delta_+)\e{-\i\phi(\theta_\text{h}^\uparrow)}\\
-u(\Delta_+)\e{-\i\phi(\theta_\text{h}^\uparrow)}\\
u(\Delta_+)\\
\end{pmatrix}\e{\i q_\text{h}^\uparrow\cos\theta_\text{h}^\uparrow x}
+ 
\frac{t_\text{h}^\downarrow}{\sqrt{2}}\begin{pmatrix}
v(\Delta_-)\\
-v(\Delta_-)\e{-\i\phi(\theta_\text{h}^\downarrow)}\\
u(\Delta_-)\e{-\i\phi(\theta_\text{h}^\downarrow)}\\
u(\Delta_-)\\
\end{pmatrix}\e{\i q_\text{h}^\downarrow\cos\theta_\text{h}^\downarrow x}.
\end{align}
\end{widetext}
We have defined $\Delta_\pm = \Delta_\text{s}\pm (\Delta_\text{t}/2)$, and
the spreading angles in Eq. (\ref{eq:wavefunctions2}) are given as $\sin \theta_e^\sigma = k_F\sin\theta/q^\sigma$, $\theta_h^\sigma = \pi - \theta_e^\sigma$. This follows from the fact that translational symmetry is conserved along the $y$-axis. The coherence factors entering the wavefunctions in Eq. (\ref{eq:wavefunctions2}) are given as
\begin{align}
u(\Delta) = \sqrt{\frac{1}{2} + \frac{\sqrt{E^2-|\Delta|^2}}{2E}},\; v(\Delta) = \sqrt{\frac{1}{2} - \frac{\sqrt{E^2-|\Delta|^2}}{2E}}.
\end{align}
We also define the dimensionless parameters $Z = 2mV_0/k_\text{F}$ and $\beta = 2m\lambda/k_\text{F}$ as a measure of the intrinsic barrier strength and magnitude of the spin-orbit coupling, respectively. 

Note that we are using the same effective masses in the normal part of the system as in the superconducting part. The mass of the quasiparticles in heavy-fermion materials are, as the name itself implies, ordinarily much larger than in normal metals. It was recently shown by Yokoyama \etal \cite{yokoyama2006} that in a two-dimensional electron gas (2DEG)/superconductor junction where spin-orbit coupling was substantial in the 2DEG, the effect of including a larger effective mass in the 2DEG was equivalent to that caused by an increase of $Z$. Note that in the presence of a time-reversal breaking magnetic field, it was shown in Refs.~\onlinecite{zutic99,zutic00} that the effect of Fermi-vector mismatch could not be reproduced simply by varying the barrier parameter $Z$. Since there is no time-reversal breaking field present in this case, however, we here restrict ourselves to considering equal effective masses in the two systems. With the above equations, one is able to find explicit expressions for $\{r_e^\sigma, r_h^\sigma\}$. The procedure illustrated here is identical for incoming electrons with $\sigma=\downarrow$ when using 
\begin{equation}
\psi(x) = 
\begin{pmatrix}
 r_e^\uparrow\e{-\i k_\text{F}\cos\theta x}\\
  \e{\i k_\text{F}\cos\theta x} + r_e^\downarrow\e{-\i k_\text{F}\cos\theta x}\\
   r_h^\uparrow\e{\i k_\text{F}\cos\theta x}\\
    r_h^\downarrow\e{\i k_\text{F}\cos\theta x}
   \end{pmatrix}
\end{equation}
instead of Eq. (\ref{eq:wavefunctions1}). This establishes the framework which serves as the basis for calculating the conductance spectrum.

\subsection{Conductance spectra for noncentrosymmetric superconductors}\label{sec:concen}
We now proceed to calculate the tunneling conductance for our setup. Generalizing the theory of Blonder, Tinkham, and Klapjwik \cite{btk}, one obtains a conductance $G(E,\theta)$ (scaled on the conductance in a N-N junction) for an incoming electron with angle $\theta$ to the junction normal with spin $\sigma$, where
\begin{align}\label{eq:cond}
G(E,\theta) &= 1 + \sum_\alpha (|r_h^\alpha(E,\theta)|^2 - |r_e^\alpha(E,\theta)|^2),
\end{align}
and $R_{N-N} = \int^{\pi/2}_{-\pi/2} [4\cos^3\theta/(4\cos^2\theta + Z^2)] \text{d}\theta$. The angularly averaged conductance reads
\begin{align}
G(E) &= (R_{N-N})^{-1} \int^{\pi/2}_{-\pi/2}  G(E,\theta) P(\theta)\text{d}\theta,
\end{align}
where $P(\theta)$ is the probability distribution function $[P(0)=1]$ for incoming electrons at an angle $\theta$. This is in many cases set to $P(\theta) = \cos\theta$, but other forms modelling \eg effective tunneling cones may also be applied. In obtaining the total conductance, one has to find $G(E)$ for both $\sigma=\uparrow$ and $\sigma=\downarrow$ and then add these contributions. The original derivation of this specific formula for the tunneling conductance given in Ref.~\onlinecite{btk} relies on the relation 
\begin{equation}\label{eq:sym}
|r_h^\sigma(E)|^2 = |r_h^\sigma(-E)|^2
\end{equation}
to hold. This is known to be valid for subgap energies, but for energies above the gap the relationship does not hold in general, a fact which implies that the conductance formula derived in Ref.~\onlinecite{btk} is only valid for applied voltages below the gap, strictly speaking. However, since the probability for Andreev reflection rapidly diminishes for energies above the gap (especially for $Z\neq0$), the conductance formula may still be applied for larger voltages as a reasonable approximation, even for the high-transparency case of low values for $Z$. 
\par
\begin{figure}[h!]
\centering
\resizebox{0.44\textwidth}{!}{
\includegraphics{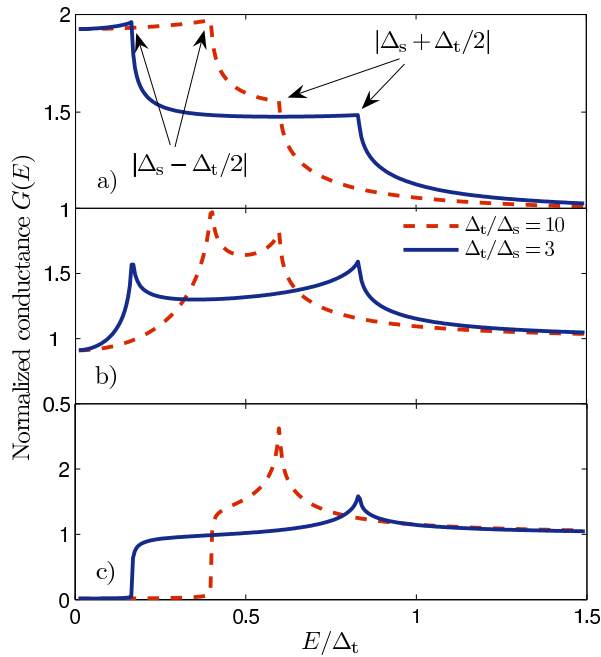}}
\caption{(Color online) Tunneling conductance for N/CePt$_3$Si junction with $\beta=0.05$. We study barrier strengths corresponding to a) $Z=0.1$, b) $Z=1$, c) $Z=10$. In all cases, we plot the ratios $\Delta_\text{t}/\Delta_\text{s} = \{3,10\}$ to see how the spectra are affected. It is seen that the conductance spectra reveal information about the relative size of the singlet and triplet component of the gaps by characteristic features located at bias voltages $E = |\Delta_\text{s} \pm \Delta_\text{t}/2|$.}
\label{fig:condNSOC}
\end{figure}

The explicit analytical expressions for the normal- and Andreev-reflection probabilities, $|r_e^\sigma|^2$ and $|r_h^\sigma|^2$ respectively, are too large and unwieldy to be of any instructive use. We shall therefore be content with plotting these expressions to reveal the physics embedded within them. In most scanning tunneling microscopy (STM) experiments, a high transparancy interface is often realized, corresponding to low $Z$. Also, since the band-splitting $2\lambda k_F$ at Fermi level may be of order \cite{samokhin2} 100 meV for CePt$_3$Si, a simple analysis relating this to our dimensionless parameter $\beta$ yields that $\beta \simeq 0.05$. We therefore plot in Fig. \ref{fig:condNSOC} the angularly averaged (and normalized) conductance spectrum for several values of barrier strength and singlet/triplet gap ratios, fixing the spin-orbit coupling parameter at $\beta=0.05$. From Fig. \ref{fig:condNSOC}, we see that one may infer the relative size of the singlet and triplet components of the gap by the characteristic behaviour of $G(E)$ at voltages corresponding to $\Delta_s \pm \Delta_t/2$. This is in agreement with what one could expect by studying the form of the eigenvalues in Eq. (\ref{eq:eigenvalues}), since it is this precise combination of the gaps that appear in the expression.
\par 
In a recent study \cite{iniotakis} by Iniotakis \etal, a normal/noncentrosymmetric superconductor junction was studied for low-transparency interfaces, where it was found that zero-bias anomalies would take place for certain STM measurement orientations if a specific form of the mixed singlet-triplet order parameter was realized. This may be attributed to the formation of zero energy bound states \cite{hu1994}, which is possible when the gap contain nodes. In the present study, we are using an isotropic spin-singlet gap and and also isotropic $p$-wave gaps $(|\Delta_{\vk\sigma\sigma}| = \text{constant})$ \cite{wang-maki}, such one does not expect the appearance of a ZBCP, in contrast to Ref.~\onlinecite{iniotakis}. Moreover, we note that the spin-orbit coupling in the system gives rise to effectively spin-active boundary conditions [see Eq. (\ref{eq:boundary})] \cite{kashiwaya, linderPRB07}. 

\par
\begin{figure}[h!]
\centering
\resizebox{0.5\textwidth}{!}{
\includegraphics{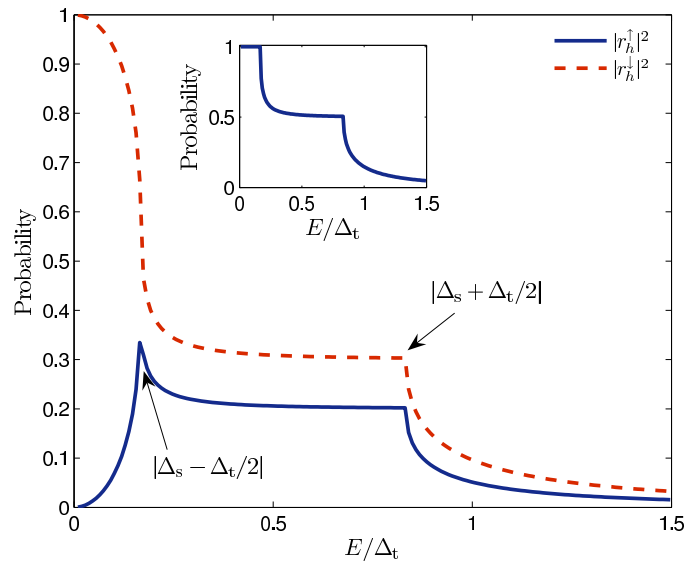}}
\caption{(Color online) Andreev-reflection coefficients for spin-$\uparrow$ and spin-$\downarrow$ fermions in the case of incoming $\sigma=\uparrow$ electrons. It is seen that the degree of spin-polarization of the generated quasiparticle current will vary with the bias voltage. The inset contains a plot of the sum of reflection coefficients (both normal and Andreev), showing that no transmittance of quasiparticles occurs for voltages below $E = |\Delta_\text{s} - \Delta_\text{t}/2|$. }
\label{fig:koeff}
\end{figure}

It is also instructive to consider the Andreev-reflection probabilities explicitly to resolve the spin-structure of the quasiparticle current, as shown in Fig. \ref{fig:koeff} for an incoming electron with spin $\sigma=\uparrow$. It is seen that the spin-$\downarrow$ coefficient becomes larger with increasing voltage, such that the spin-polarization of the current will vary with the bias voltage. The proper definition of a spin-current in systems exhibiting spin-orbit coupling has, however, been shown \cite{shi} to be more subtle than applying the usual relations for charge- and spin-currents
\begin{equation}
\mathbf{j}_\text{charge} = -e\sum_\sigma \mathbf{j}_\sigma,\; \mathbf{j}_\text{spin} = \sum_\sigma \sigma\mathbf{j}_\sigma,
\end{equation}
where $\mathbf{j}_\sigma$ is the particle-current of fermions with spin $\sigma$.
Therefore, it is fair to claim that it is not obvious how one might detect such a change in polarization of the quasiparticle current with a change in bias voltage. On the other hand, the charge-current remains unaffected by these considerations and our results thus indicate that the conductance spectrum of the charge-current in a N/CePt$_3$Si junction may provide valuable information about the relative size of the singlet and triplet components of the superconductivity gap.

We now comment on effects that have not been taken into account in our analysis of this 
problem. First, the issue of how boundary effects affect the order parameters is addressed. 
Studies \cite{ambegaokar1974, buchholtz1981, tanuma2001} have shown that interfaces/surfaces may
have a pair-breaking effect on unconventional superconductivity order parameters. This is 
relevant in tunneling junction experiments as in the present case. The suppression of the order 
parameter is caused by a formation of so-called midgap surface states (also known as zero-energy states) 
\cite{hu1994} which occurs for certain orientations of the $\vk$-dependent superconducting gaps that 
satisfy a resonance condition. Note that this is not the case for conventional $s$-wave
superconductors since the gap is isotropic in that case. This pair-breaking surface effect was 
studied specifically for $p$-wave order parameters in Refs.~\onlinecite{ambegaokar1974, buchholtz1981}, and it was found that the component of the order parameter that experiences a sign change under the transformation $k_\perp \to -k_\perp$, where $k_\perp$ is the component of momentum perpendicular to the tunneling junction, was suppressed in the vicinity of the junction. By vicinity of the junction, we here mean a distance comparable to the coherence length, typically of order 1-10 nm. Thus, depending on the explicit form of the superconducting gaps in the noncentrosymmetric superconductor, these could be suppressed close to the interface. Moreover, we are dealing with an easily 
observable effect, since distinguishing between the peaks 
occuring for various values of $R_\Delta$ requires a resolution 
of order ${\cal{O}}(10^{-1}\Delta_{\uparrow,0})$, which typically corresponds to $0.1-1$ meV. These structures should readily be resolved with present-day STM technology. However,  it should be pointed out that a challenge with respect to tunneling junctions is dealing with non-idealities at the interface which may affect the conductance spectrum. \\
\indent In order to fully consider the possible pair-breaking effect of the interface in an enhanced model, one would obviously need to solve the scattering problem self-consistently in order to obtain more precise results for the conductance, especially in terms of the quantitative aspect. To obtain analytical results, however, we have inserted the gaps \textit{a priori}, since we believe that our model captures essential qualitative features in a N/CePt$_3$Si junction that could be probed for. This belief is motivated by studies \cite{suppression} for $d_{x^2-y^2}$ superconductors which show that the conductance shape around zero bias remains essentially unchanged even if the spatial dependence of the order parameters are taken into account. The spectra around the gap edges may be modified in the sense that since the gap in general will be somewhat reduced close the interface, the appearance of characteristic features in the conductance could occur at lower bias voltages than the bulk value of the gaps. However, it seems reasonable to hope that our simple model may be of use in predicting qualitative features of the conductance spectrum when considering junctions involving noncentrosymmetric superconductors such as CePt$_3$Si.

\section{Probing the pairing symmetry of ferromagnetic superconductors}\label{sec:fermag}
As a second application of our model, we consider a model of a ferromagnetic superconductor described by uniformly coexisting itinerant ferromagnetism and equal-spin pairing non-unitary spin-triplet superconductivity. 

\subsection{Model and formulation}
We write down a mean-field theory Hamiltonian with equal-spin pairing Cooper pairs and a finite magnetization along the easy-axis similar to the model studied in Refs.~\onlinecite{nevidomskyy, gronsleth, bedell}, namely
\begin{align}
\hat{H} &= \sum_\vk \xi_\vk + \frac{INM^2}{2} - \frac{1}{2}\sum_{\vk\sigma} \Delta_{\vk\sigma\sigma}^\dag b_{\vk\sigma\sigma} \notag\\
&+\frac{1}{2}\sum_{\vk\sigma} \Big(\hat{c}_{\vk\sigma}^\dag \hat{c}_{-\vk\sigma}\Big)
\begin{pmatrix}
\xi_{\vk\sigma} & \Delta_{\vk\sigma\sigma} \\
\Delta_{\vk\sigma\sigma}^\dag & -\xi_{\vk\sigma} \\
\end{pmatrix}
\begin{pmatrix}
\cop_{\vk\sigma}\\
\cop_{-\vk\sigma}^\dag\\
\end{pmatrix},
\end{align}
Applying the diagonalization procedure described in Sec. \ref{sec:model}, we arrive at 
\begin{align}
\hat{H} &= H_0 + \sum_{\vk\sigma} E_{\vk\sigma} \hat{\gamma}_{\vk\sigma}^\dag\hat{\gamma}_{\vk\sigma},\notag\\
H_0 &= \frac{1}{2}\sum_{\vk\sigma}(\xi_{\vk\sigma} - E_{\vk\sigma} - \Delta_{\vk\sigma\sigma}^\dag b_{\vk\sigma\sigma}) + \frac{INM^2}{2},
\end{align}
where $\{\hat{\gamma}_{\vk\sigma},\hat{\gamma}_{\vk\sigma}^\dag\}$ are new fermion operators and the eigenvalues read
\begin{equation}
E_{\vk\sigma} = \sqrt{\xi_{\vk\sigma}^2 + |\Delta_{\vk\sigma\sigma}|^2}.
\end{equation}
Recall that it is implicit in our notation that $\xi_\vk$ is measured from Fermi level. The free energy is obtained by using the procedure explained in Sec. \ref{sec:model}, and one obtains
\begin{align}
F = H_0 - \frac{1}{\beta}\sum_{\vk\sigma}\text{ln}(1 + \e{-\beta E_{\vk\sigma}}),
\end{align}
such that the gap equations for the magnetic and superconducting order parameters become \cite{nevidomskyy}
\begin{align}
M = -\frac{1}{N} \sum_{\vk\sigma} \frac{\sigma \xi_{\vk\sigma}}{2 E_{\vk\sigma}} \text{tanh}(\beta E_{\vk\sigma}/2),\notag\\
\Delta_{\vk\sigma\sigma} = -\frac{1}{N}\sum_{\vk'} V_{\vk\vk'\sigma\sigma} \frac{\Delta_{\vk'\sigma\sigma}}{2 E_{\vk'\sigma}}\text{tanh}(\beta E_{\vk'\sigma}/2).
\end{align}
For concreteness, we now consider a specific form of the gaps, similar to those studied in Refs.~\onlinecite{nevidomskyy, bedell}. Assuming that the gap is fixed on the Fermi surface in the weak-coupling limit, we write
\begin{equation}\label{eq:gaptriplet}
\Delta_{\vk\sigma\sigma} = \Delta_{\bar{\vk}_F\sigma\sigma} = \frac{\Delta_{\sigma,0}}{\sqrt{3/8\pi}}Y^\sigma_{l=1}(\theta,\phi),
\end{equation}
where $\bar{\vk}_F$ is the normalized Fermi wave-vector, such that the gap only depends on the direction of the latter. We have introduced the spherical harmonics
\begin{equation}\label{eq:Y}
Y^\sigma_{l=1}(\theta,\phi) = -\sigma\sqrt{3/8\pi}\e{\i\sigma\theta}\sin\phi,
\end{equation}
such that the gaps in Eq. (\ref{eq:gaptriplet}) experience a change in sign under inversion of momentum, \ie $\theta \to \theta + \pi$. We shall consider the case $\sin\phi = 1$ which renders the magnitude of the gaps to be constant, similar to the A2-phase in liquid $^3$He. The motivation for this is that it seems plausible that uniform coexistence of ferromagnetic and superconducting order may only be realized in thin-film structures where the Meissner (diamagnetic) response of the superconductor is suppressed for in-plane magnetic fields. This enables us to set $\sin\phi = 1$, since the electrons are restricted from moving in the $\hat{\mathbf{z}}$-direction in a thin-film structure. In a bulk structure, as considered in Ref.~\onlinecite{bedell}, we expect that a spontaneous vortex lattice should be the favored thermodynamical state \cite{tewari2004}. The pairing potential may then in general be written as
\begin{align}
V_{\sigma\sigma}(\theta,\theta') = -\sum_m \frac{g^m_{\sigma\sigma}}{3/8\pi} Y^\sigma(\theta)[Y^\sigma(\theta')]^*,
\end{align}
which for the chosen gaps reduces to 
\begin{equation}
V_{\sigma\sigma}(\theta,\theta') = -\frac{8\pi g}{3}Y^\sigma(\theta)[Y^\sigma(\theta')]^*.
\end{equation}
Conversion to integral gap equations is accomplished by means of the identity
\begin{equation}
\frac{1}{N} \sum_\vk f(\xi_{\vk\sigma}) = \int \text{d}\varepsilon N^\sigma(\varepsilon),
\end{equation}
where $N^\sigma(\varepsilon)$ is the spin-resolved density of states. In three spatial dimensions, this may be calculated from the dispersion relation by using the formula
\begin{equation}
N^\sigma(\varepsilon) = \frac{V}{(2\pi)^3} \int_{\varepsilon_{\vk\sigma} = \text{const}} \frac{\text{d} S_{\varepsilon_{\vk\sigma}}}{|\hat{\nabla}_\vk \varepsilon_{\vk\sigma}|}.
\end{equation}
With the dispersion relation $\xi_{\vk\sigma}= \varepsilon_\vk - \sigma IM - E_F$ (having set the chemical potential equal to the Fermi energy, $\mu=E_F$), one obtains
\begin{equation}
N^\sigma(\varepsilon) = \frac{mV\sqrt{2m(\varepsilon + \sigma IM + E_F)}}{2\pi^2}.
\end{equation}
In their integral form, the gap equations read
\begin{align}\label{eq:gapeqint}
M &= -\frac{1}{2}\sum_\sigma \sigma \int_{-E_F-\sigma IM}^{\infty} \text{d}\varepsilon \frac{\varepsilon N^\sigma(\varepsilon)}{\sqrt{\varepsilon^2 + \Delta_{\sigma,0}^2}}\text{tanh}[\beta E_\sigma(\varepsilon)/2],\notag\\
1 &= \frac{g}{2} \int^{\omega_0}_{-\omega_0} \text{d}\varepsilon \frac{ N^\sigma(\varepsilon) }{E_\sigma(\varepsilon)}\text{tanh}[\beta E_\sigma(\varepsilon)/2].
\end{align}

\subsection{Zero temperature case}
\begin{figure}[h!]
\centering
\resizebox{0.5\textwidth}{!}{
\includegraphics{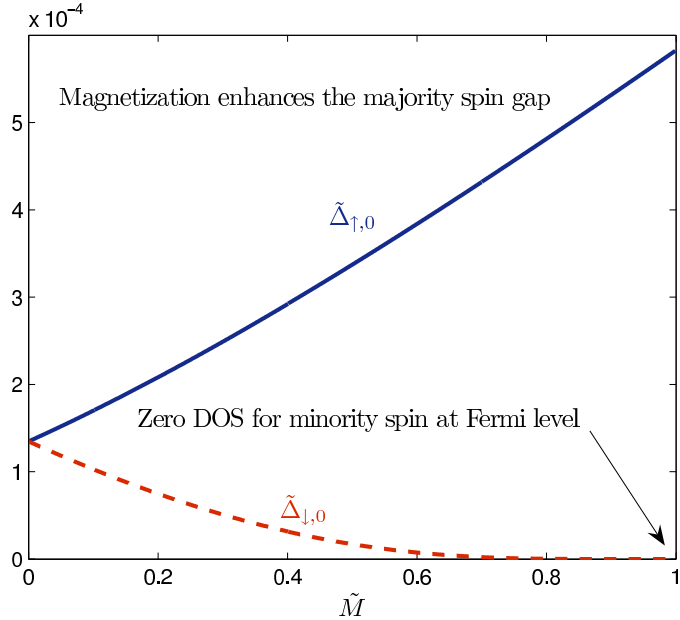}}
\caption{(Color online) Superconducting gaps (full drawn line: majority spin, dashed line: minority spin) as a function of the magnetization with $\omega_0 = 0.01E_F$. When the exchange splitting equals the Fermi energy, the DOS of minority spin fermions is zero at Fermi level, resulting in a complete suppression of $\Delta_{\sigma,0}$.}
\label{fig:gapsM}
\end{figure}

Consider now $T=0$, where we are able to obtain analytical expressions for the superconductivity order parameters in the problem. Since the superconductivity gap equation reduces to
\begin{equation}
1 = \frac{g}{2} \int^{\omega_0}_{-\omega_0} \text{d}\varepsilon \frac{ N^\sigma(\varepsilon) }{E_\sigma(\varepsilon)},
\end{equation}
one readily finds
\begin{equation}\label{eq:gapanalytical}
\Delta_{\sigma,0} = 2\omega_0 \e{-1/c\sqrt{1+\sigma\tilde{M}}},\; \sigma=\uparrow,\downarrow
\end{equation}
where we have defined $\tilde{M} = IM/E_F$, \ie the exchange energy scaled on the Fermi energy. Moreover, the weak coupling constant $c = gN(0)/2$ will be set to 0.2 throughout the rest of this paper, unless specifically stated otherwise. Moreover, we set $\tilde{\omega}_0 = \omega_0/E_F = 0.01$ as the typical spectral width of the bosons responsible for the attractive pairing potential. From Eq. (\ref{eq:gapanalytical}), we see that the effect of increasing the magnetization is an increase in the gap for majority spin. The important influence of the magnetization is that it modifies the density of states, which affects the superconductivity gaps. For $\tilde{M} = 1$, \ie an exchange splitting equal to the Fermi energy, the minority spin gap is completely suppressed, as shown in Fig. \ref{fig:gapsM}. Thus, the presence of magnetization reduces the available phase space for the minority spin Cooper pairs, suppressing the gap and the critical temperature compared to the pure Bardeen-Cooper-Schrieffer (BCS) case. 

\begin{figure}[h!]
\centering
\resizebox{0.5\textwidth}{!}{
\includegraphics{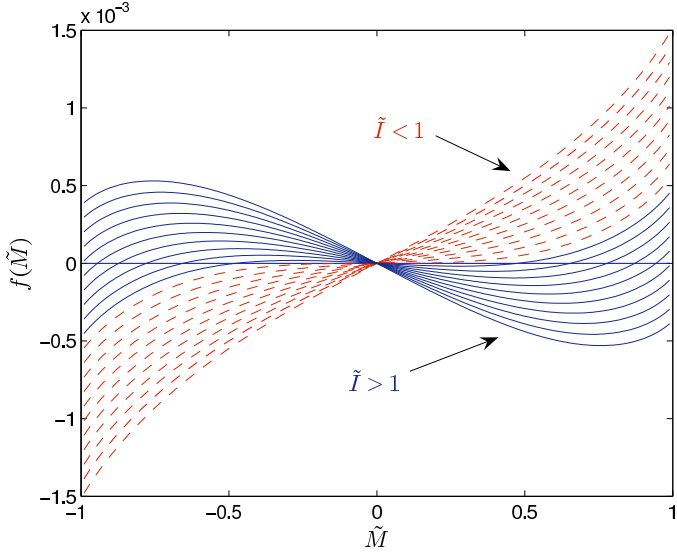}}
\caption{(Color online) Plot of $f(\tilde{M})$ in Eq. (\ref{eq:gapM}) as a function of $\tilde{M}$. Only the trivial solution exists for $\tilde{I}<1$, while three possible solutions are present at $\tilde{I} > 1$. We have plotted Eq. (\ref{eq:gapM}) for values of $\tilde{I} = [0.9, 1.1]$ in steps of $0.01$.}
\label{fig:plotM}
\end{figure}

After the appropriate algebraic manipulations of Eq. (\ref{eq:gapeqint}), the self-consistency equation for the magnetization becomes
\begin{align}\label{eq:gapM}
f(\tilde{M}) &= \tilde{M} +\frac{\tilde{I}}{4} \int^\infty_{-1-\tilde{M}} \text{d}x\sqrt{1+x+\tilde{M}}\notag\\
&\times\Bigg\{\frac{\Big[1-2\Theta\Big(-\sqrt{x^2 + \tilde{\Delta}_{\uparrow,0}^2(\tilde{M})}\Big)\Big]}{x^{-1}\sqrt{x^2 + \tilde{\Delta}_{\uparrow,0}^2(\tilde{M})}} \notag\\
&- \frac{\Big[1-2\Theta\Big(-\sqrt{(x+2\tilde{M})^2 + \tilde{\Delta}_{\downarrow,0}^2(\tilde{M})}\Big)\Big]}{(x+2\tilde{M})^{-1}\sqrt{(x+2\tilde{M})^2 + \tilde{\Delta}_{\downarrow,0}^2(\tilde{M})}}\Bigg\} = 0,
\end{align}
where we have defined the parameter $\tilde{I} = IN(0)$, in similarity to Ref.~\onlinecite{nevidomskyy}, and introduced
$\tilde{\Delta}_{\sigma,0}(\tilde{M}) = \Delta_{\sigma,0}/E_F$.
We have thus managed to decouple the gap equations completely, such that one only has to solve Eq. (\ref{eq:gapM}) to find the magnetization, and then plug that value into Eq. (\ref{eq:gapanalytical}). Note that strictly speaking, one should divide the integral in Eq. (\ref{eq:gapM}) into three parts: $\{-1-\tilde{M}, -\omega_0\}, \{\omega_0,\infty\}, \{-\omega_0,\omega_0\}$ where the superconductivity gaps are only non-zero in the latter interval. However, the error associated with doing the integration numerically over the entire regime with a finite value for the gaps is completely neglible. From Eq. (\ref{eq:gapM}), we see that the trivial solution $\tilde{M}=0$ is always possible. Interestingly, we find that a non-trivial solution implying coexistence of ferromagnetism and superconductivity is only possible when $\tilde{I} > 1$ (in agreement with Ref.~\onlinecite{nevidomskyy}). To illustrate this fact, consider Fig. \ref{fig:plotM} for a plot of $f(\tilde{M})$ in Eq. (\ref{eq:gapM}) as a function of $\tilde{M}$ for several values of $\tilde{I}$. In fact, it is seen that more than one solution is possible for any $\tilde{I} > 1$: the trivial solution $\tilde{M}=0$ corresponding to a unitary superconducting state, and a non-trivial solution $M = \tilde{M}_0$, representing a non-unitary superconducting state. Recall that in terms of the $\mathbf{d}_\vk$-vector formalism, these classifications are defined as
\begin{equation}
\text{Unitary:}\; \mathbf{d}_\vk\times\mathbf{d}_\vk^* = 0,\;\; \text{Non-unitary:}\; \mathbf{d}_\vk\times\mathbf{d}_\vk^* \neq 0.
\end{equation}
We will later show that the free energy is minimal in the non-unitary state, which implies that the coexistence of ferromagnetism and superconductivity may indeed be realized in our model. 

\begin{figure}[h!]
\centering
\resizebox{0.5\textwidth}{!}{
\includegraphics{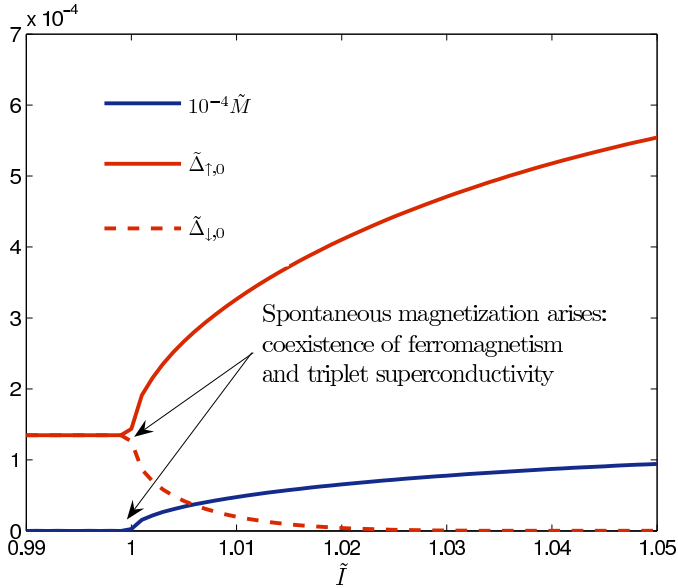}}
\caption{(Color online) Self-consistently solved order parameters at $T=0$ as a function of $\tilde{I}$. It is seen that the coexistent regime of ferromagnetism and superconductivity is located at $\tilde{I}>1$, where a spontaneous magnetization arises.}
\label{fig:OPzeroT}
\end{figure}

\indent The order parameters depend on the parameters $(T,\tilde{I},c)$. To illustrate their dependence on $\tilde{I}$ at $T=0$, consider Fig. \ref{fig:OPzeroT}. It is clearly seen that the superconductivity gaps are equal for $\tilde{I}<1$, corresponding to a unitary spin-triplet pairing state. For $\tilde{I}>1$, a spontaneous magnetization arises and the majority/minority spin gap increases/decreases. This corresponds to the coexistent phase of ferromagnetism and superconductivity. An important point concerning Eq. (\ref{eq:gapM}) is the inclusion of the step-function factors, which are superfluous as long as we are considering the coexistent regime of ferromagnetism and superconductivity, since their argument is always negative. However, if one for instance were to set one or both of the superconductivity gaps to zero, the correct gap equation for the magnetization would not be reproduced without them. This is due to the loss of generality in taking the limit $\text{tanh}(\beta E_\sigma)\to1$ when $\beta\to\infty$ in deriving Eq. (\ref{eq:gapM}), since $E_\sigma>0$ is replaced with $\varepsilon$ when superconductivity is lost, which can be both larger and smaller than zero when $\Delta_{\sigma,0}\to 0$. The present form of Eq. (\ref{eq:gapM}) is generally valid for the purely magnetic and the coexistent A1- and A2-phases of the ferromagnetic superconductor.

\begin{figure}[h!]
\centering
\resizebox{0.5\textwidth}{!}{
\includegraphics{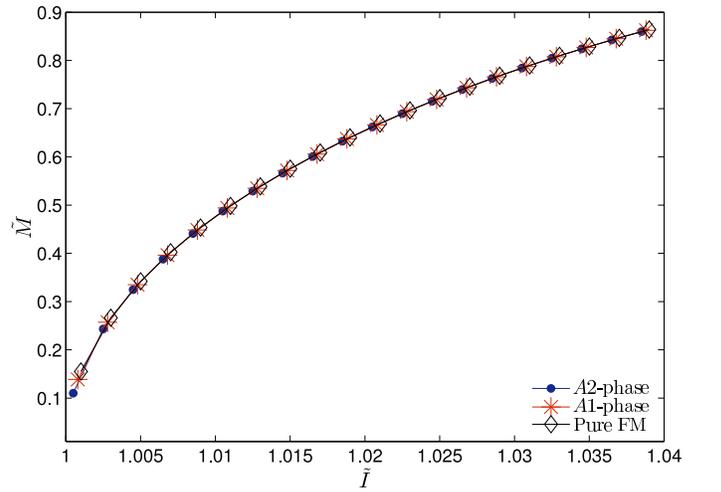}}
\caption{(Color online) Plot of the exchange energy $\tilde{M} = IM/E_F$ at $T=0$ as a function of $\tilde{I} = IN(0)$ for two possible phases of a FMSC (the $A1$- and $A2$-phase) compared to the purely ferromagnetic case. It is seen that $\tilde{M}$ is virtually unaltered by the presence of superconductivity, at least in the weak-coupling approach we have adopted here (see also Ref.~\onlinecite{bedell}).}
\label{fig:MT0sfaI}
\end{figure}

In order to correctly characterize the pairing symmetry of FMSCs, it is of interest to find clear-cut experimental signatures that distinguish between the possible phases of such an unconventional superconductors. As we have alluded to, it seems reasonable to assume that a superconducting phase analogous to the $A1$- or $A2$-phase of $^3$He may be realized in FMSCs. We now investigate how the magnetization at $T=0$ depends on the ferromagnetic exchange energy constant $\tilde{I}$ in these possible phases, and compare them to the purely ferromagnetic case. Our results are shown in Fig. \ref{fig:MT0sfaI}, where we have self-consistently solved for $\tilde{M}$ as a function of $\tilde{I}$ in three cases: 1) the purely ferromagnetic phase, 2) the $A1$-phase where only spin-$\uparrow$ fermions are paired, and 3) the $A2$-phase where all spin-bands participate in the superconducting pairing. It is seen that the magnetization is practically identical in all phases regardless of the value of $\tilde{I}$. Analytically, this may be understood since the difference $\Delta f$ [see Eq. (\ref{eq:gapM})] between the gap equation for the magnetization in the purely ferromagnetic case and the coexistent state reads
\begin{align}
\Delta f = \sum_\sigma \sigma \Bigg[ \int^{\omega_0}_{-\omega_0}\text{d}\varepsilon N^\sigma(\varepsilon) \Big(1-\frac{|\varepsilon|}{\sqrt{\varepsilon^2+\Delta_{\sigma,0}^2}}\Big) \Bigg] \simeq 0.
\end{align}
Note that in our results, an enhancement of the magnetization below the superconductivity critical temperature is absent, contrary to the results of Ref.~\onlinecite{bedell} who predicted that the magnetization should be enhanced in the coexistent phases compared to the purely ferromagnetic phase. For the weak-coupling approach applied here, it seems reasonable that the presence of superconductivity should not alter the magnetization much, while superconductivity itself is drastically modified depending on the strength of the exchange energy. The result of Ref.~\onlinecite{bedell} may be a consequence of the fact that they do not set $\sin\phi = 1$ [Eq. (\ref{eq:Y})], and consequently have additional nodes compared to the gaps we are using. 

\subsection{Finite temperature case}
\begin{figure}[h!]
\centering
\resizebox{0.5\textwidth}{!}{
\includegraphics{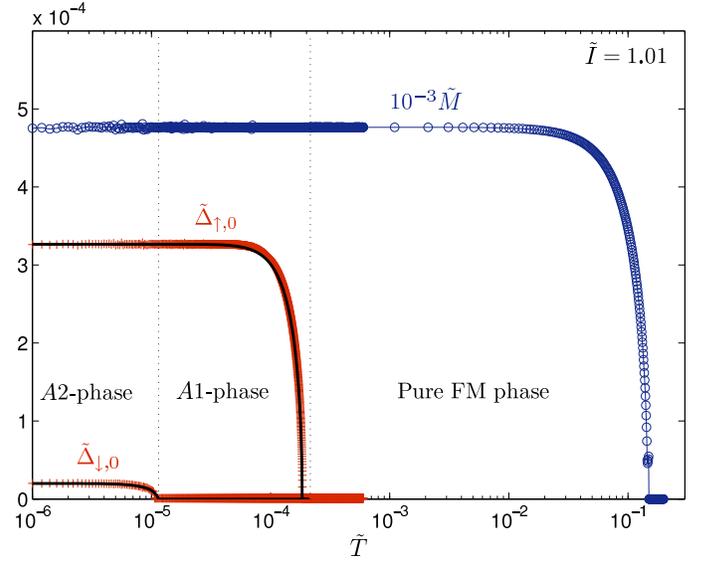}}
\caption{(Color online) Self-consistently solved order parameters as a function of temperature for $\tilde{I} = 1.01$. Note that the temperature axis is logarithmic, such that the transition between the paramagnetic and ferromagnetic phase is much higher than the superconducting phase transitions. However, we are able to tune $\tilde{I}$ such that $T_{c,M}$ and $T_{c,\sigma}$ become comparable. We have also plotted the gaps with self-consistently solved values at $T=0$ and then applying a BCS-temperature dependence (solid black lines), which yield excellent consistency with the solution that does not assume a BCS-temperature dependence.  }
\label{fig:I101}
\end{figure}

The critical temperature for the superconductivity order parameter is found by solving the equation
\begin{equation}
1 = \frac{g}{2}\int^{\omega_0}_{-\omega_0}\text{d}\varepsilon \frac{N^\sigma(\varepsilon)}{\varepsilon}\text{tanh}(\varepsilon/2T_{c,\sigma}),
\end{equation}
which yields the BCS-like solution
\begin{equation}\label{eq:crit}
T_{c,\sigma} = 1.13\omega_0\e{-1/c\sqrt{1+\sigma\tilde{M(T_{c,\sigma})}}}.
\end{equation}
Since the transition temperature for paramagnetism - ferromagnetism is in general much larger than the superconducting phase transition, one may to good approximation set $M(T_{c,\sigma}) = M(0)$.
It is then evident that the critical temperature depends on the magnetization in the same manner as the gap itself, and the cutoff-dependence in Eq. (\ref{eq:gapanalytical}) may be removed in favor of the critical temperature by substituting Eq. (\ref{eq:crit}). In order to solve the coupled gap equations self-consistently at arbitrary temperature, we considered Eq. (\ref{eq:gapeqint}) with the result given in Fig. \ref{fig:I101}. It is seen that the minority-spin gap is clearly suppressed compared to the majority-spin gap in the presence of a net magnetization. Also, the graph clearly shows that the BCS-temperature dependence constitutes an excellent approximation for the decrease of the OPs with temperature. In what follows, we shall therefore use self-consistently obtained solutions at $T=0$ for the OPs and make use of the BCS temperature-dependence unless specifically stated otherwise. In general, the critical temperature for the ferromagnetic order parameter, $T_{c,M}$ exceeds the superconducting phase transition temperatures $T_{c,\sigma}$ by several orders of magnitude. However, for $\tilde{I}$ very close to one, we are able to make these transition temperatures comparable in magnitude. In the experimentally discovered FMSCs UGe$_2$ and URhGe, one finds that $T_{c,M}$ is 50-100 times higher than the temperature at which superconductivity arises. 

\begin{figure}[h!]
\centering
\resizebox{0.5\textwidth}{!}{
\includegraphics{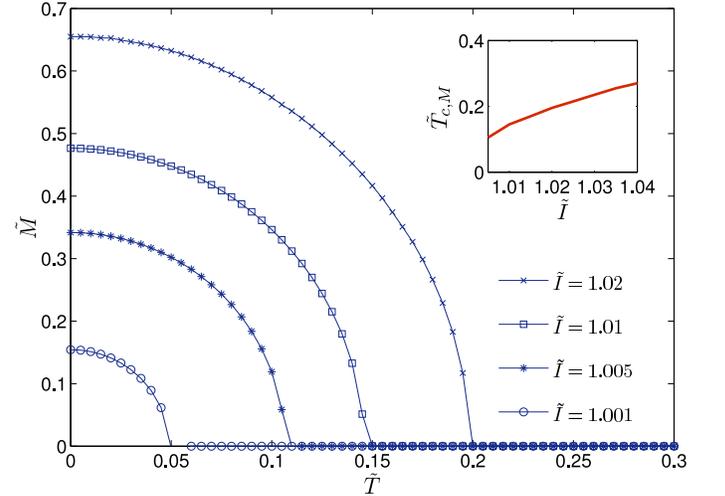}}
\caption{(Color online) Temperature-dependence of the magnetic order parameter for several values of $\tilde{I}$. The insets shows how the critical temperature depends on $\tilde{I}$.}
\label{fig:KritiskMTc}
\end{figure}

To illustrate how the magnetic order parameter depends on $\tilde{I}$, consider Fig. \ref{fig:KritiskMTc} for a plot of the temperature dependence for several values of $\tilde{I}$. The inset shows how the critical temperature depends on this parameter.

\subsection{Comparison of free energies}
\begin{figure}[h!]
\centering
\resizebox{0.5\textwidth}{!}{
\includegraphics{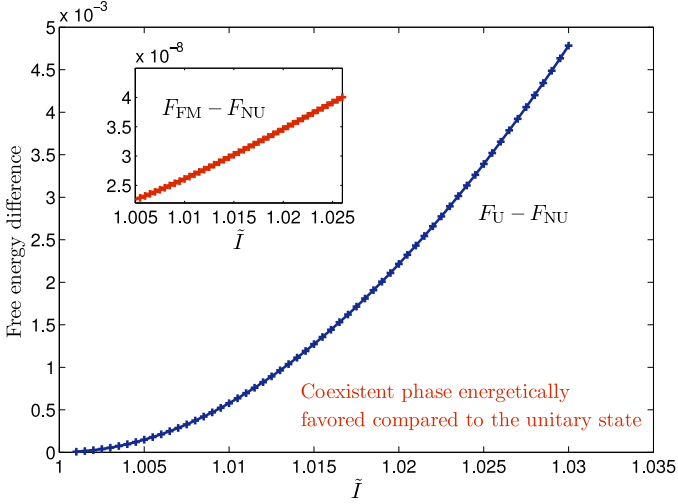}}
\caption{(Color online) Comparison between the free energy for the non-unitary and unitary superconducting state at zero temperature. It is seen that these values are equal for $\tilde{I}=1$ (and for $\tilde{I}<1$), while the non-unitary state is energetically favored for increasing ferromagnetic exchange energy. Thus, the coexistent phase should be realized at sufficiently low temperatures in the presence of a ferromagnetic exchange energy. }
\label{fig:freeenergy}
\end{figure}

Although a non-trivial solution of $M$ exists, care must be exercised before concluding that this is the preferred energetical configuration of the system. Specifically, it may in theory be possible that the systems prefers the $M=0$ solution regardless of the value of $\tilde{I}$, corresponding to a unitary superconducting state with $\Delta_{\uparrow,0}=\Delta_{\downarrow,0}$. It is therefore necessary to compare the free energies of the $M=0$ and $M\neq0$ cases at values of $\tilde{I}$ where the latter is a possible solution, and also study their temperature dependence. In the general case, the analytical expression for the free energy in the coexistent non-unitary superconducting phase reads
\begin{align}
F/N &= \frac{IM^2}{2} + \sum_\sigma \frac{\Delta_{\sigma,0}^2}{2g} - \sum_\sigma  \int_{-E_F-\sigma IM}^\infty \text{d}\varepsilon N^\sigma(\varepsilon)\notag\\
&\times\Bigg[\frac{\sqrt{\varepsilon^2 + \Delta_{\sigma,0}^2}}{2} + \frac{1}{\beta}\text{ln}(1 + \e{-\beta\sqrt{\varepsilon^2 + \Delta_{\sigma,0}^2}})\Bigg].
\end{align}
Note that the gap should be set to zero in the above equation everywhere except in the interval $\{-\omega_0,\omega_0\}$. We obtain a dimensionless measure of the free energy by multiplying with $I/E_F^2$, and denote $F_\text{NU} = FI/(NE_F^2)$.
Note that the free energies of the unitary state, pure ferromagnetic state, and paramagnetic state are obtained as follows:
\begin{align}
F_\text{U} &= \lim_{M\to0} F_\text{NU},\notag\\
F_\text{PM} &= \lim_{M\to0, \Delta_{\sigma,0}\to 0} F_\text{NU},\notag\\
F_\text{FM} &= \lim_{\Delta_{\sigma,0}\to0} F_\text{NU}.
\end{align}
In Fig. \ref{fig:freeenergy}, we plot the difference between the unitary and non-unitary solution at zero temperature, $\Delta F = F_\text{U}-F_\text{NU}$, which clearly shows how the system favors the non-unitary solution with spontaneous magnetization as $\tilde{I}$ increases. As a result, we suggest that the coexistent phase of ferromagnetism and superconductivity should be realized at sufficiently low temperatures whenever a magnetic exchange energy is present. For consistency, we also verified that $F_\text{NU} < F_\text{FM}$ at $T=0$ since the system otherwise would prefer to leave superconductivity out of the picture and stay purely ferromagnetic.

\begin{figure}[h!]
\centering
\resizebox{0.5\textwidth}{!}{
\includegraphics{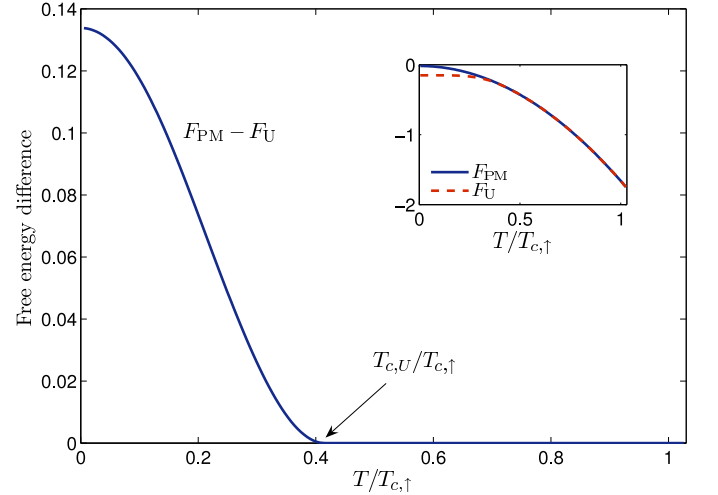}}
\caption{(Color online) Free energy difference between the paramagnetic state ($F_\text{PM}$) and the unitary superconducting state ($F_\text{U}$). In consistency with established results (see \eg Ref.~\onlinecite{tinkham}), the free energies merge continuously as the temperature gets closer to $T_{c,U}$. In the inset, we have chosen the zero-temperature value of the paramagnetic free energy as zero, serving as a reference point. We have solved all order parameters self-consistently for $\tilde{I}=1.01$.}
\label{fig:PMU}
\end{figure}

\begin{figure}[h!]
\centering
\resizebox{0.5\textwidth}{!}{
\includegraphics{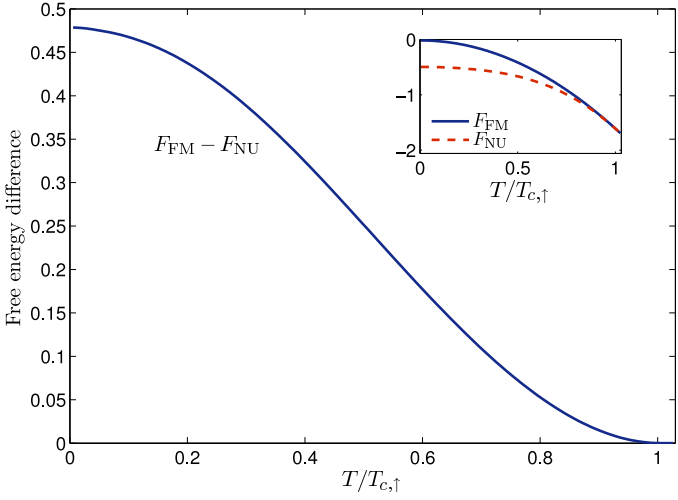}}
\caption{(Color online) Free energy difference between the ferromagnetic state ($F_\text{FM}$) and the non-unitary superconducting state ($F_\text{NU}$) which displays coexistence of ferromagnetism and superconductivity. It is seen that the non-unitary phase is favored compared to the purely ferromagnetic state. In the inset, we have chosen the zero-temperature value of the ferromagnetic free energy as zero, serving as a reference point. We have solved all order parameters self-consistently for $\tilde{I}=1.01$. The curves of Figs. 12 and 13 may be made congruent by 
a simple scaling of the axes. This is a consequence of the weak-coupling limit, where superconductivity
sets in at a temperature much smaller than the ferromagnet-paramagnet transition temperature, such the that
magnetic order parameter across the superconducting transition essentially is a temperature-independent
constant.}
\label{fig:FMNU}
\end{figure}

\begin{figure}[h!]
\centering
\resizebox{0.5\textwidth}{!}{
\includegraphics{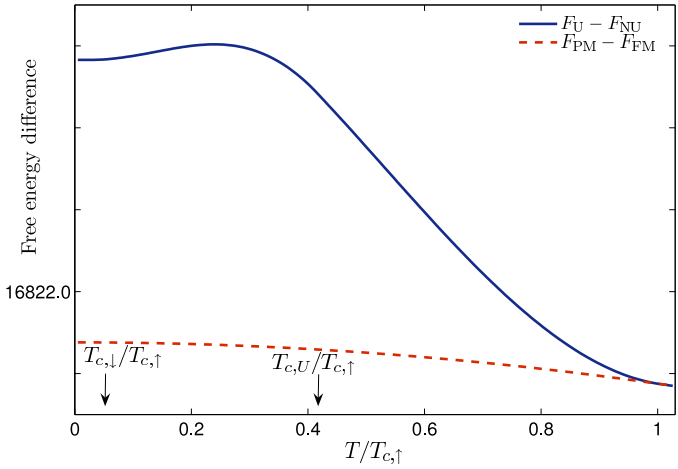}}
\caption{(Color online) Free energy difference between the unitary and non-unitary state ($F_\text{U} - F_\text{NU}$) as well as the paramagnetic and ferromagnetic state ($F_\text{PM} - F_\text{FM}$). At $T=T_{c,\uparrow}$, the curves merge smoothly into each other since all superconductivity is lost. Each step along the ordinate corresponds to an increment of unit 0.1. We have solved all order parameters self-consistently for $\tilde{I}=1.01$.}
\label{fig:Alle4}
\end{figure}

We now turn to the temperature-dependence of the free energy at the fixed value of $\tilde{I}=1.01$ (the order parameters were self-consistently solved for this value and plotted in Fig. \ref{fig:I101}). The results are shown in Figs. \ref{fig:PMU} to \ref{fig:Alle4}. Note that we now use a different scaling of the free energy, namely $F_\text{NU} = F/[NN(0)T_{c,\uparrow}^2]$. The well-known result that the free energy of a purely superconducting state joins the free energy of the paramagnetic state continuously as the temperature increases is reproduced in Fig. \ref{fig:PMU}. In Fig. \ref{fig:FMNU}, we see that the coexistent phase of ferromagnetism and superconductivity is energetically favored compared to the purely ferromagnetic case, which is consistent with the experimental fact that a transition to superconductivity occurs below the Curie temperature for certain materials \cite{aoki, saxena}. Finally, in Fig. \ref{fig:Alle4}, we have plotted the energy difference between the unitary and non-unitary free energy in addition to the difference between the paramagnetic and ferromagnetic phases. It is seen that the non-unitary state is energetically preferred over the unitary state, a statement which strictly speaking has only been shown to hold for our current choice of $\tilde{I}$ ($\tilde{I}=1.01$), but it seems reasonable to assume that it holds under quite general circumstances due to the presence of an exchange energy. At $T=T_{c,\uparrow}$, when all superconductivity is lost, the two curves join each other smoothly since $F_\text{NU}\to F_\text{FM}$ and $F_\text{U}\to F_\text{PM}$ when $T>T_{c,\uparrow}$. Our results then suggest the very real possibility of a coexistent phase of spin-triplet superconducting pairing and itinerant ferromagnetism being realized in the experimentally discovered ferromagnetic superconductors, since we have shown that the coexistent phase is energetically favored over both the purely magnetic and the non-magnetic superconducting state.

\subsection{Specific heat}
We next consider some experimental signatures that could be expected in the different possible phases of a FMSC. Consequently, we have calculated the electronic contribution to the specific heat of the system by making use of $C_V =  T\frac{\partial S}{\partial T}$ with
\begin{equation}
S = -\sum_{\vk\sigma} \{ f(E_{\vk\sigma})\text{ln}[f(E_{\vk\sigma})] + [1-f(E_{\vk\sigma})]\text{ln}[1 - f(E_{\vk\sigma})]\}
\end{equation}
as the entropy, leading to
\begin{align}
C_V = \frac{\beta^2}{4} \sum_{\vk\sigma} \frac{E_{\vk\sigma}^2 - \beta^{-1}(\Delta_{\sigma,0}\frac{\partial \Delta_{\sigma,0}}{\partial T} - \sigma\varepsilon_{\vk\sigma}I\frac{\partial M}{\partial T})}{\text{cosh}^2(\beta E_{\vk\sigma}/2)}.
\end{align}
Note that the above equation reduces to the correct normal-state heat capacity in the limit $\{\Delta_{\sigma,0}, M\}\to0$, with the usual linear $T$-dependence. The term $\frac{\partial \Delta_{\sigma,0}}{\partial T}$ ensures that the well-known mean-field BCS discontinuity (strictly speaking valid only for a type-I superconductor
\cite{HLS1974}, but 
clearly invalid
at the transition temperature of a strong type II superconductor \cite{DH1981,tesanovic1999,nguyen-sudbo1999}) at the superconducting critical temperature is present in the heat capacity, while the presence of ferromagnetism induces a new term proportional to $\frac{\partial M}{\partial T}$. However, due to our previous argument that $T_{c,M} \gg T_{c,\sigma}$, this term may be neglected since the magnetization remains virtually unaltered in the temperature regime around $T_{c,\sigma}$. Going to the integral representation of the equation for the heat capacity, one thus obtains
\begin{align}\label{eq:CV}
C_V = \frac{\beta^2}{4}\sum_\sigma \int_{-E_F-\sigma IM}^\infty \text{d}&\varepsilon N^\sigma(\varepsilon) [E^2_\sigma(\varepsilon) - \frac{\partial \Delta_{\sigma,0}}{\partial T}\Delta_{\sigma,0}T]\notag\\
&\times\text{cosh}^{-2}[\beta E_\sigma(\varepsilon)/2].
\end{align}
Strictly speaking, one should again divide the above integral into the regions $\{-E_F,-\omega_0\}$, $\{\omega_0,\infty\}$, and $\{-\omega_0,\omega_0\}$ where the superconductivity gap should be set to zero in all regions except the latter. However, since the integrand is strongly peaked around $\varepsilon = 0$ (Fermi level), there is little error made in using the form Eq. (\ref{eq:CV}). In order to obtain the derivatives of the gap functions with respect to temperature, an analytical approach is permissable since the gaps have the BCS-temperature dependence (see Fig. \ref{fig:I101})
\begin{align}
\Delta_{\sigma,0}(T) = \Delta_{\sigma,0}(0)\text{tanh}\Big(1.74\sqrt{T_{c,\sigma}/T - 1}\Big),
\end{align}
where the superconductivity critical temperature for spin-$\sigma$ fermions is given by Eq. (\ref{eq:crit}). To illustrate how the superconductivity pairing symmetry leaves important fingerprints in the heat capacity, we solved Eq. (\ref{eq:CV}) self-consistently for two values of $\tilde{I}$ corresponding to strong ($\tilde{M} \simeq 0.5$) and weak ($\tilde{M} \simeq 0.1$) exchange splitting. At $\tilde{I} = 1.01$, the discontinuity is clearly pronounced for $T=T_{c,\uparrow}$, but it is hardly discernable at $T=T_{c,\downarrow}$. However, for $\tilde{I}=1.0005$ where the superconductivity transition temperatures for majority and minority spins become comparable, a clear double-peak signature is revealed in the heat capacity. We thus propose that this particular feature should serve as unambigous evidence of a superconducting pairing corresponding to the $A2$-phase of liquid $^3$He in ferromagnetic superconductors.

\begin{figure}[h!]
\centering
\resizebox{0.5\textwidth}{!}{
\includegraphics{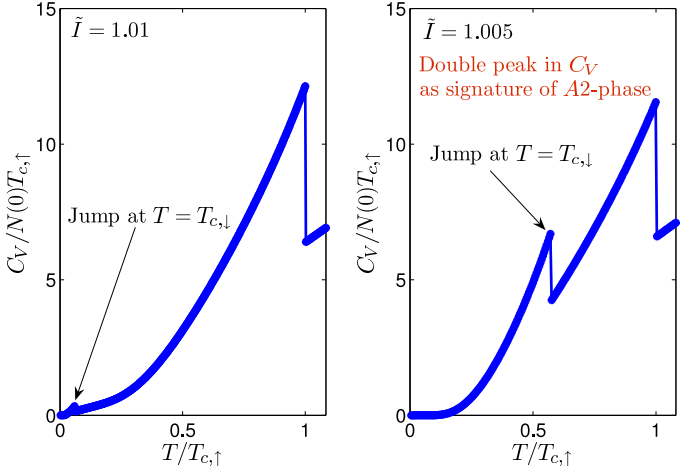}}
\caption{(Color online) Specific heat capacity as a function of temperature for two values of $\tilde{I}$, corresponding to a strong exchange splitting ($\tilde{M} \simeq 0.5$) and a weak exchange splitting ($\tilde{M} \simeq 0.1$). A double-peak signature is clearly visible when the transition temperatures for the majority and minority spin bands are comparable. }
\label{fig:CV}
\end{figure}

\begin{figure}[h!]
\centering
\resizebox{0.5\textwidth}{!}{
\includegraphics{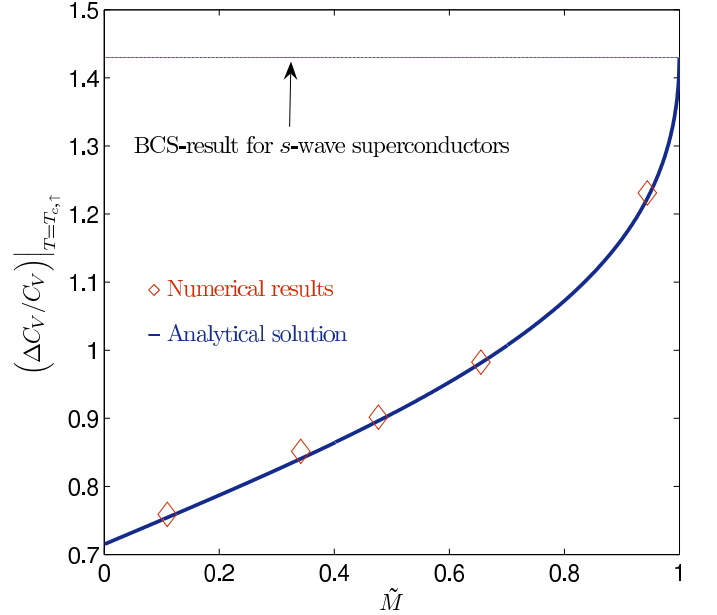}}
\caption{(Color online) The discontinuity of the heat capacity at $T=T_{c,\uparrow}$ as a function of exchange splitting [Eq. (\ref{eq:jump})]. It is seen that the BCS value is recovered at $\tilde{M}=1$. Note that it would also be recovered at $\tilde{M}=0$, although this is not shown explicitely in the figure. The reason for this is that we have assumed that $T_{c,\downarrow}\neq T_{c,\uparrow}$. We have also plotted the numerical results ($\diamond$) for the jump with self-consistently solved OPs, \ie without assuming BCS temperature dependence, for $\tilde{I} = \{1.001,1.005,1.01,1.02,1.05\}$, which yield good agreement with the analytical solution Eq. (\ref{eq:jump}).}
\label{fig:discont}
\end{figure}

An classic feature of the BCS-theory of superconductivity was the prediction that the jump in the heat capacity at $T_c$ normalized on the normal-state value was a universal number, namely 
\begin{equation}
\Big( \frac{\Delta C_V}{C_V} \Big) \Big|_{T=T_c} \simeq 1.43.
\end{equation}
In the presence of a net magnetization, one would expect that the universality of this ratio would break down and depend on the strength of the exchange energy. 
This is due to the fact that the discontinuity in the specific heat at the 
superconducting transition  is dominated by the majority-spin carriers, while the total 
specific heat to a larger extent has contributions from both minority-spin and majority spin carriers. 
To investigate this statement quantitatively, we consider the jump in $C_V$ at $T=T_{c,\uparrow}$ since no analytical approach is possible at $T=T_{c,\downarrow}$, as seen from Eq. (\ref{eq:CV}). We find that the normal (ferromagnetic) state heat capacity reads
\begin{equation}
C_V^\text{FM} = \frac{\pi^2 T_{c,\uparrow}}{3}\sum_\sigma N^\sigma(0),
\end{equation}
where $N^\sigma(0)$ is the spin-resolved DOS at Fermi level, while the difference between the heat capacity in the coexistent state and the ferromagnetic state at $T=T_{c,\uparrow}$ reads
\begin{equation}
\Delta C_V = \frac{1.74^2\Delta_{\uparrow,0}^2(0)N^\uparrow(0)}{2T_{c,\uparrow}}.
\end{equation}
Since the zero-temperature value for the gap is $\Delta_{\uparrow,0}(0) = 1.76T_{c,\uparrow}$, one arrives at
\begin{align}\label{eq:jump}
\Big( \frac{\Delta C_V}{C_V} \Big) \Big|_{T=T_{c,\uparrow}} = 1.43 \frac{1}{1 + \sqrt{\frac{1-\tilde{M}}{1+\tilde{M}}}}.
\end{align}
The above equation reduces to the BCS-limit for complete spin-polarization $\tilde{M}=1$ (zero DOS for spin-$\downarrow$ fermions at Fermi level). 
This is due to, as noted above, the larger extent to which majority-spin carriers dominate the
{\it jump} in specific heat compared to the total specific heat. 
As anticipated, the jump in $C_V$ depends on the exchange energy, as illustrated in Fig. \ref{fig:discont}. Of course, in the unitary state $\tilde{M}=0$ the jump also reduces to the BCS value although this is not seen from Eq. (\ref{eq:jump}). The reason for this is that we have implicitly assumed that $\tilde{M}\neq0$ in the derivation of Eq. (\ref{eq:jump}), taking $T_{c,\uparrow} > T_{c,\downarrow}$. In the case where these transition temperatures are equal, the contribution from both is additive and equal [1.43/2, to be specific, as seen from Eq. (\ref{eq:jump})] and gives the correct BCS result.

Our study of $C_V$ then offers two interesting opportunities: \textit{i)} the presence or absence of a double-peak signature in the heat capacity reveals information about the superconductivity pairing symmetry realized in the FMSC, and \textit{ii)} the normalized value of the discontinuous jump at $T_{c,\uparrow}$ contains information about the exchange splitting between the majority and minority spin carrier bands. 

\section{Summary}\label{sec:summary}
In summary, we have derived a general Hamiltonian describing coexistence of itinerant ferromagnetism,  spin-orbit coupling and mixed spin-singlet/triplet superconducting pairing using mean-field theory. Exact eigenvalues and coupled gap equations for the different order parameters have been obtained. Our results may serve as a starting point for any model describing coexistence of any combination of these three phenomena simply by applying the appropriate limit. \\
\indent As a specific application of our results, we have studied quantum transport between a normal metal and a superconductor lacking an inversion center with mixed singlet and triplet gaps. We find that there are pronounced peaks and bumps in the conductance spectrum at voltages corresponding to the sum and difference of the magnitude of the singlet and triplet gaps. Consequently, our results may be helpful in obtaining information about the size of the relative contribution of different pairing symmetries. \\
\indent Moreover, we considered a system where itinerant ferromagnetism uniformly coexists with spin-triplet superconductivity as a second application of our theory. We solved the coupled gap equations numerically, and presented analytical expressions for the order parameters and their dependences on quantities such as exchange energy and temperature. It was found that the coexistent regime of ferromagnetism and superconductivity may indeed be realized since it is energetically favored compared to a unitary superconducting state ($M=0$) and a purely ferromagnetic state. In order to make contact with the experimental situation, we studied the heat capacity and found interesting signatures in the spectrum that may be used in order to obtain information about both the superconductivity pairing symmetry present in the system and the magnitude of the exchange energy. \\

\section{Acknowledgments}
J. L. gratefully acknowledges A. Nevidomskyy for stimulating communications, and thanks E. K. Dahl, S. Kragset, and K. Berland for providing useful comments. This work was
  supported by the Norwegian Research Council Grants No. 157798/432
  and No. 158547/431 (NANOMAT), and Grant No. 167498/V30 (STORFORSK).
  
\appendix

\section{Bogoliubov-de Gennes equations for systems exhibiting coexistence of ferromagnetism, spin-orbit coupling, and superconductivity}
\subsection{Derivation}
We start out with a real-space Hamiltonian described by fermionic field operators $\psi^{(\dag)}(\vecr,t)$ with a general attractive pairing kernel $V_{\alpha\beta}(|\vecr-\vecr'|)$, namely
\begin{align}
\hat{H} &= \sum_{\alpha\beta} \int \text{d}\vecr \psi^\dag_\alpha(\vecr,t)\Big[ - \frac{\hat{\nabla}_\vecr^2}{2m} - \mu + \{V+\alpha V_s)\delta(x) \notag\\
&\hspace{0.9in} + [-\vV_M + \vg(\hat{\vp})]\cdot\check{\boldsymbol{\sigma}}\Big]_{\alpha\beta} \psi^\dag_\beta(\vecr,t)\notag\\
&+ \frac{1}{2}\sum_{\alpha\beta} \int \int \text{d}\vecr\text{d}\vecr' V_{\alpha\beta}(|\vecr-\vecr'|) \psi_\alpha^\dag(\vecr,t)\psi^\dag_\beta(\vecr',t)\notag\\
&\hspace{0.7in}\times\psi_\beta(\vecr',t)\psi_\alpha(\vecr,t).
\end{align}
Here, $V_0$ accounts for a non-magnetic scattering potential associated with a barrier located at $x=0$ while $V_s$ is the magnetic scattering potential, \ie the barrier is spin-active. Moreover, $\vV_M$ is the magnetic exchange energy vector, $\vg(\hat{\vp}) = - \vg(-\hat{\vp})$ is a term describing an antisymmetric spin-orbit coupling energy $(\hat{\vp} = -\i\hat{\nabla}_\vecr)$, while $\check{\boldsymbol{\sigma}}$ is the vector of Pauli matrices. We now introduce the mean-field approximation
\begin{equation}
\psi^\dag_\alpha(\vecr,t)\psi^\dag_\beta(\vecr') = \langle \psi^\dag_\alpha(\vecr,t)\psi^\dag_\beta(\vecr',t) \rangle + \delta\psi_{\alpha\beta}^\dag,
\end{equation}
where the last term describes the flucuations around the average field, and also define the superconducting order parameter
\begin{equation}
\Delta_{\alpha\beta}(\vecr,\vecr') = V_{\alpha\beta}(|\vecr-\vecr'|) \langle \psi_\beta(\vecr',t)\psi_\alpha(\vecr,t) \rangle.
\end{equation}
Above, we have explicitly made the superconductivity order parameter time-independent, which effectively amounts to saying that it does not depend on energy (the weak-coupling limit).
This provides us with 
\begin{align}
\hat{H} &= \sum_{\alpha\beta} \int \text{d}\vecr \psi^\dag_\alpha(\vecr,t)\Big[ - \frac{\hat{\nabla}_\vecr^2}{2m} - \mu + \{V+\alpha V_s)\delta(x) \notag\\
&\hspace{0.5in} + [-\vV_M + \vg(\hat{\vp})]\cdot\check{\boldsymbol{\sigma}}\Big]_{\alpha\beta} \psi^\dag_\beta(\vecr,t)\notag\\
&+ \frac{1}{2}\sum_{\alpha\beta} \int \int \text{d}\vecr\text{d}\vecr' [ \Delta_{\alpha\beta}^\dag(\vecr,\vecr')\psi_\beta(\vecr',t)\psi_\alpha(\vecr,t) \notag\\
&\hspace{0.3in} + \Delta_{\alpha\beta}(\vecr,\vecr')\psi^\dag_\alpha(\vecr,t)\psi_\beta^\dag(\vecr',t) \notag\\
&\hspace{0.1in}- V_{\alpha\beta}(|\vecr-\vecr'|) \langle\psi_\alpha^\dag(\vecr,t)\psi^\dag_\beta(\vecr',t)\rangle\langle\psi_\beta(\vecr',t)\psi_\alpha(\vecr,t)\rangle].
\end{align}
The time-dependent field operators $\psi(\vecr,t) = \e{\i \hat{H}t} \psi(\vecr) \e{-\i \hat{H}t}$ obey the Heisenberg equations of motion
\begin{align}
\i\partial_t\psi_\alpha(\vecr,t) &= [\psi_\alpha(\vecr,t), \hat{H}]\notag\\
&= \sum_\beta \int \text{d}\vecr' \delta(\vecr-\vecr') \hat{H}^0_{\alpha\beta}(\vecr',\hat{\vp}) \psi_\beta(\vecr',t) \notag\\
&+\sum_\beta \int \text{d}\vecr' \Delta_{\alpha\beta}(\vecr,\vecr') \psi_\beta^\dag(\vecr',t),\notag\\
\i\partial_t\psi^\dag_\alpha(\vecr,t) &= [\psi^\dag_\alpha(\vecr,t), \hat{H}]\notag\\
&= \sum_\beta \int \text{d}\vecr' \delta(\vecr-\vecr') [-\hat{H}^0(\vecr',-\hat{\vp})]_{\alpha\beta}^\mathcal{T} \psi^\dag_\beta(\vecr',t) \notag\\
&+\sum_\beta \int \text{d}\vecr' \Delta^\dag_{\alpha\beta}(\vecr,\vecr') \psi_\beta(\vecr',t).
\end{align}
For convenience, we have defined
\begin{align}
\hat{H}_{\alpha\beta}^0(\vecr,\hat{\vp}) &= \Big[ - \frac{\hat{\nabla}_\vecr^2}{2m} - \mu + (V+\alpha V_s)\delta(x)\notag\\
&+ [-\vV_M + \vg(\hat{\vp})]\cdot\check{\boldsymbol{\sigma}}\Big]_{\alpha\beta}.
\end{align}
The above equations may be comprised in compact matrix form
\begin{align}
\i\partial_t\Psi(\vecr,t) &= \int \text{d}\vecr' \mathcal{H}(\vecr,\vecr')\Psi(\vecr',t),\notag\\
\Psi(\vecr,t) &= [\psi_\uparrow(\vecr,t), \psi_\downarrow(\vecr,t), \psi^\dag_\uparrow(\vecr,t), \psi^\dag_\downarrow(\vecr,t)]^\text{T},\notag\\
\mathcal{H}(\vecr,\vecr') &= 
\begin{pmatrix}
\hat{H}^0(\vecr',\hat{\vp})\delta_{\vecr\vecr'} & \hat{\boldsymbol{\Delta}}(\vecr, \vecr') \\
-\hat{\boldsymbol{\Delta}}^*(\vecr, \vecr') & [-\hat{H}^0(\vecr',-\hat{\vp})]^\mathcal{T}\delta_{\vecr\vecr'}  \\
\end{pmatrix},
\end{align}
with $\delta(\vecr-\vecr') = \delta_{\vecr\vecr'}$, and where we have defined 
\begin{equation}
\hat{\boldsymbol{\Delta}}(\vecr, \vecr') = \begin{pmatrix}
\Delta_{\uparrow\uparrow}(\vecr, \vecr') & \Delta_{\uparrow\downarrow}(\vecr, \vecr') \\
\Delta_{\downarrow\uparrow}(\vecr, \vecr') & \Delta_{\downarrow\downarrow}(\vecr, \vecr') \\
\end{pmatrix}.
\end{equation}
Note that $\Delta_{\uparrow\downarrow}(\vecr, \vecr')$ is in general a superposition of a triplet (T) and singlet (S) component that satisfy
\begin{align}
\Delta_{\uparrow\downarrow}(\vecr, \vecr') &= \Delta^\text{T}_{\uparrow\downarrow}(\vecr, \vecr') + \Delta^\text{S}_{\uparrow\downarrow}(\vecr, \vecr'),\notag\\
\Delta^\text{T}_{\uparrow\downarrow}(\vecr, \vecr') &= \Delta^\text{T}_{\downarrow\uparrow}(\vecr, \vecr'),\notag\\
\Delta^\text{S}_{\uparrow\downarrow}(\vecr, \vecr') &= -\Delta^\text{S}_{\downarrow\uparrow}(\vecr, \vecr').
\end{align}
Regarding $\Psi(\vecr, t)$ as a $c$-number and assuming a stationary solution $\Psi(\vecr,t) = \Psi(\vecr) \e{-\i Et}$ with $E$ as the wavefunction energy, it suffices to solve the equation
\begin{equation}
E\Psi(\vecr) = \int \text{d}\vecr' \mathcal{H}(\vecr,\vecr') \Psi(\vecr').
\end{equation}
By considering a plane-wave solution of $\Psi(\vecr)$ and dividing out the fast oscillations on an atomic-scale (see \eg Ref.~\onlinecite{bruder}), one is left with most familiar form of the BdG-equations appearing in the literature, namely
\begin{align}\label{eq:bdgusual}
\begin{pmatrix}
\hat{H}^0(\vecr,\hat{\vp}) & \hat{\boldsymbol{\Delta}}(\vk,\vecr) \\
\hat{\boldsymbol{\Delta}}^\dag(\vk,\vecr) & [-\hat{H}^0(\vecr,-\hat{\vp})]^\mathcal{T} \\
\end{pmatrix}\Psi(\vecr) = E\Psi(\vecr),
\end{align}
where the quasiparticle momentum $\vk$ is the Fourier-transform of the relative-coordinate $\mathbf{s} = (\vecr-\vecr')/2$, \ie
\begin{equation}
\mathcal{F}\{f(\vk)\} = \int \text{d}\mathbf{s} f(\mathbf{s})\e{-\i\vk\cdot\mathbf{s}}. 
\end{equation}
This is usually assumed to be fixed on the Fermi surface, such that only the directional dependence of $\vk$ enters in Eq. (\ref{eq:bdgusual}), $\vk \to k_F \hat{\vk}$.

\subsection{Boundary conditions}

We proceed to provide a general approach in order to obtain the correct boundary conditions at the interface for the wavefunctions. Continuity of the wavefunction itself is assumed in this context. Consider our Eq. (\ref{eq:matrixnoncentro}) which describes the Hamiltonian for the N/CePt$_3$Si junction. The first row of the equation explicitly reads
\begin{align}
\Big[-&\frac{1}{2m}\frac{\partial^2}{\partial x^2} -\frac{1}{2m}\frac{\partial^2}{\partial y^2} - \mu + V_0\delta(x)\Big]\psi_\uparrow(x,y) \notag\\
&+ \lambda(\frac{\partial}{\partial x} -\i\frac{\partial}{\partial y})\Theta(x)\psi_\downarrow(x,y) +  \Delta_{\vk\uparrow\uparrow}\Theta(x) \psi_\uparrow^\dag(x,y)\notag\\
&+ \Delta_\text{s}\Theta(-x) \psi_\downarrow^\dag(x,y) = E\psi_\uparrow(x,y),
\end{align}
If we now integrate the above equation over a an interval $[\epsilon, -\epsilon]$ along the $\hat{\mathbf{x}}$-axis and apply the limit $\epsilon\to0^+$, one obtains
\begin{align}
\lim_{\epsilon\to0^+}\Big\{-\frac{1}{2m} &[\psi'_\uparrow(\epsilon,y) - \psi'_\uparrow(-\epsilon,y)]  + V_0\psi_\uparrow(0,y) \notag\\
&+ \lambda \int^\epsilon_{-\epsilon} \text{d}x [\Theta(x)\psi_\downarrow(x,y)]'\Big\}= 0,
\end{align}
where $'$ denotes derivation with respect to $x$. The last term yields $\frac{1}{2}\lambda\psi_\downarrow(\epsilon,0)$ (since $\Theta(0) = \frac{1}{2}$), such that the boundary condition for derivative of the $\psi_\uparrow(x,y)$-component becomes
\begin{align}
\lim_{\epsilon\to0^+}\Big\{[&\psi'_\uparrow(\epsilon,y) - \psi'_\uparrow(-\epsilon,y)] - m\lambda\psi_\downarrow(\epsilon,0) \Big\} \notag\\
&= 2mV_0\psi_\uparrow(0,y).
\end{align}
It is seen that the presence of spin-orbit coupling and the delta-function barrier leads to a discontinuity of the derivative of the wave-function. A similar procedure may be applied to the other components of $\Psi(x,y)$, and this method can also be extended to include different effective masses on each side of the junction modelled by a simple step-function $\Theta(x)$.


\begin{thebibliography}{99}

\bibitem{saxena} S. S. Saxena , P. Agarwal, K. Ahilan, F. M. Grosche, R. K. W. Haselwimmer, M. J. Steiner, E. Pugh, I. R. Walker, S. R. Julian, P. Monthoux, G. G. Lonzarich, A. Huxley, I. Sheikin, D. Braithwaite, and J. Flouquet, Nature {\bf 406}, 587 (2000).

\bibitem{aoki}   D. Aoki, A. Huxley, E. Ressouche, D. Braithwaite, J. Flouquet, J.-P. Brison, E. Lhotel, and C. Paulsen, Nature {\bf 413}, 613 (2001).

\bibitem{bauer1} E. Bauer, G. Hilscher, H. Michor, Ch. Paul, E. W. Scheidt, A. Gribanov, Yu. Seropegin, H. Noël, M. Sigrist, and P. Rogl , Phys. Rev. Lett. \textbf{92}, 027003 (2004).
\bibitem{akazawa1} T. Akazawa, H. Hidaka, T. Fujiwara, T. C. Kobayashi, E. Yamamoto, Y. Haga, R. Settai, and Y. Onuki,  
, J. Phys. Cond. Mat. \textbf{16}, L29 (2004).

\bibitem{huxley} F. Hardy, A. D. Huxley,  
Phys. Rev. Lett. {\bf 94}, 247006 (2005).

\bibitem{samokhin} K. V. Samokhin, M. B. Walker, 
Phys. Rev. B {\bf 66}, 174501 (2002).

\bibitem{machida} K. Machida, T. Ohmi, Phys. Rev. Lett. 
{\bf 86}, 850 (2001).

\bibitem{nelson} K. D. Nelson, Z. Q. Mao, Y. Maeno, 
and Y. Liu, Science {\bf 306}, 1151 (2004).

\bibitem{sergienko} I. A. Sergienko, V. Keppens, M. McGuire, R. Jin, J. He, 
S. H. Curnoe, B. C. Sales, P. Blaha, D. J. Singh, K. Schwarz, and D. Mandrus, 
Phys. Rev. Lett. \textbf{92}, 065501 (2004).

\bibitem{yuan} H. Q. Yuan, D. F. Agterberg, N. Hayashi, P. Badica, D. Vandervelde, 
K. Togano, M. Sigrist, and M. B. Salamon, Phys. Rev. Lett. {\bf 97}, 017006 (2006).

\bibitem{curro} N. J. Curro {\it et al.}, Nature {\bf 434}, 622 (2005).

\bibitem{lebed} A. G. Lebed, Phys. Rev. Lett. {\bf 96}, 037002 (2006).

\bibitem{mazin} I. Zutic and I. Mazin, Phys. Rev. Lett. {\bf 95}, 217004 (2005).

\bibitem{andersonbook} P. W. Anderson, \textit{Basic Notions of Condensed Matter Physics}, Addison
Wesley (1980).

\bibitem{yogi} M. Yogi, Y. Kitaoka, S. Hashimoto, T. Yasuda, R. Settai, T. D. Matsuda, Y. Haga, 
Y. Onuki, P. Rogl, and E. Bauer, Phys. Rev. Lett. \textbf{93}, 027003 (2004).

\bibitem{edelstein} V. M. Edelstein, Sov. Phys. JETP \textbf{68}, 1244 (1989); V. M. Edelstein, 
Phys. Rev. Lett. \textbf{75}, 2004 (1995).

\bibitem{gorkov} L. P. Gor'kov and E. I. Rashba, Phys. Rev. Lett. \textbf{87}, 037004 (2001).

\bibitem{sergienko2} I. A. Sergienko and S. H. Curnoe, Phys. Rev. B \textbf{70}, 214510 (2004).

\bibitem{borkje} K. B{\o}rkje and A. Sudb{\o}, Phys. Rev. B \textbf{74}, 054506 (2006).

\bibitem{frigeri2} P. A. Frigeri, D.F. Agterberg, I. Milat, M. Sigrist, cond-mat/0505108.

\bibitem{frigeri} P. Frigeri, D. F. Agterberg, A. Koga, and M. Sigrist, 
Phys. Rev. Lett. \textbf{92}, 097001 (2004).

\bibitem{izawa} K. Izawa, Y. Kasahara, Y. Matsuda, K. Behnia, T. Yasuda, R. Settai, 
and Y. Onuki,  Phys. Rev. Lett. \textbf{94}, 197002 (2005).

\bibitem{tanaka2} T. Yokoyama, Y. Tanaka, and J. Inoue, 
Phys. Rev. B \textbf{72}, 220504(R) (2005).

\bibitem{anderson} P. W. Anderson, Phys. Rev. B \textbf{30}, 4000 (1984).

\bibitem{eremin2} I. Eremin and J. F. Annett, Phys. Rev. B \textbf{74}, 184524 (2006).

\bibitem{samokhin2} K. V. Samokhin, E. S. Zijlstra, and S K. Bose, 
Phys. Rev. B \textbf{69}, 094514 (2004).

\bibitem{niu} J. Shi and Q. Niu, cond-mat/0601531. 

\bibitem{tewari2004} S. Tewari, D. Belitz, T. R. Kirkpatrick, and J. Toner, 
Phys. Rev. Lett. \textbf{93}, 177002 (2004).

\bibitem{shopova2005} D. V. Shopova and D. I. Uzunov, Phys. 
Rev. B \textbf{72}, 024531 (2005).

\bibitem{mineev2005} V. P. Mineev, cond-mat/0507572.

\bibitem{mineev1999} V. P. Mineev and K. V. Samokhin, 
Introduction to Unconventional Superconductivity 
(Gordon and Breach, New York, 1999).

\bibitem{kotegawa2005} H. Kotegawa, A. Harada, S. Kawasaki, 
Y. Kawasaki, Y. Kitaoka, Y. Haga, E. Yamamoto, Y. Onuki, 
K. M. Itoh, and E. E. Haller, J. Phys. Soc. Jpn. \textbf{74}, 
705 (2005).

\bibitem{kulic2005} M. L. Kulic, C. R. Physique \textbf{7}, 
4 (2006); M. L. Kulic, and I. M. Kulic,  Phys. Rev. B 
\textbf{63}, 104503 (2001).

\bibitem{eremin2006} I. Eremin, F. S. Nogueira, and R.-J. Tarento, 
Phys. Rev. B \textbf{73}, 054507 (2006).

\bibitem{hardy2005} F. Hardy and A. D. Huxley, 
Phys. Rev. Lett. \textbf{94}, 247006 (2005).

\bibitem{samokhin2002} K. V. Samokhin and M. B. Walker, 
Phys. Rev. B \textbf{66}, 174501 (2002).

\bibitem{linder} J. Linder, M. Gr{\o}nsleth, A. Sudb{\o}, 
Phys. Rev. B \textbf{75}, 054518 (2007).

\bibitem{yokoyama07} T. Yokoyama and Y. Tanaka, 
Phys. Rev. B \textbf{75}, 132503 (2007).

 \bibitem{leggett1975} A. J. Leggett, Rev. Mod. Phys. \textbf{47}, 331 (1975).

 \bibitem{edwards} C. H. Edwards, Jr., D. E. Penney, \textit{Elementary Linear Algebra}, Prentice Hall (1988).

 \bibitem{abramowitz} M. Abramowitz and I. A. Stegun, \textit{Handbook of Mathematical Functions}, Dover, New York (1972).

\bibitem{molenkamp} L. W. Molenkamp, G. Schmidt, and G. E. W. Bauer, Phys. Rev. B \textbf{64}, 121202(R) (2001).
 
\bibitem{tanaka} Y. Tanaka and S. Kashiwaya, Phys. Rev. Lett. \textbf{74}, 3451 (1995).

\bibitem{tanaka97}  Y. Tanaka, S. Kashiwaya, Phys. Rev. B \textbf{56}, 892 (1997).

\bibitem{yokoyama2006} T. Yokoyama, Y. Tanaka, and J. Inoue, 
Phys. Rev. B \textbf{74}, 035318 (2006).

\bibitem{zutic99} I. Zutic and O. T. Valls, Phys. Rev. B \textbf{60}, 6320 (1999).

\bibitem{zutic00} I. Zutic and O. T. Valls, Phys. Rev. B \textbf{61}, 1555 (2000).

\bibitem{btk} G. E. Blonder, M. Tinkham, and T. M. Klapwijk, Phys. Rev. B \textbf{25}, 4515 (1982).

\bibitem{iniotakis} C. Iniotakis, N. Hayashi, Y. Sawa, T. Yokoyama, U. May, Y. Tanaka, M. Sigrist, cond-mat/0701643.

\bibitem{kashiwaya} S. Kashiwaya, Y. Tanaka, N. Yoshida, and M. R. Beasley, Phys. Rev. B \textbf{60}, 3572 (1999).

\bibitem{linderPRB07} J. Linder and A. Sudb{\o}, Phys. Rev. B \textbf{75}, 134509 (2007).

\bibitem{shi} J. Shi, P. Zhang, D. Xiao, and Q. Niu, Phys. Rev. Lett. \textbf{96}, 076604 (2006) 

\bibitem{ambegaokar1974} V. Ambegaokar, P. G. deGennes, and D. Rainer, Phys.
Rev. A \textbf{9}, 2676 (1974).
\bibitem{buchholtz1981} L. J. Buchholtz and G. Zwicknagl, Phys. Rev. B \textbf{23}, 5788 (1981).

\bibitem{tanuma2001} Y. Tanuma, Y. Tanaka, S. Kashiwaya, Phys. Rev. B \textbf{64}, 214519 (2001).

\bibitem{hu1994} C.-R. Hu, Phys. Rev. Lett. \textbf{72}, 1526 (1995).

\bibitem{wang-maki} G. F. Wang and K. Maki, Europhys. Lett. \textbf{45}, 71 (1999).


\bibitem{suppression} Y. S. Barash, H. Burkhardt, D. Rainer, Phys. Rev. Lett. \textbf{77}, 4070 (1996); 
Y. Tanaka and S. Kashiwaya, Phys. Rev. B \textbf{58}, 2948 (1998); 
Y. Tanaka, T. Asai, N. Yoshida, J. Inoue, and S. Kashiwaya , Phys. Rev. B \textbf{61}, R11902 (2000).

\bibitem{nevidomskyy} A. H. Nevidomskyy, Phys. Rev. Lett. \textbf{94}, 097003 (2005).

\bibitem{gronsleth} M. S. Gr{\o}nsleth, J. Linder, J.-M. B{\o}rven, and A. Sudb{\o}, Phys. Rev. Lett. \textbf{97}, 147002 (2006); J. Linder, M. S. Gr{\o}nsleth, A. Sudb{\o}, Phys. Rev. B \textbf{75}, 024508 (2007).

\bibitem{bedell} H. P. Dahal, J. Jackiewicz, K. S. Bedell, Phys. Rev. B \textbf{72}, 172506 (2005).

\bibitem{tinkham} M. Tinkham, \textit{Introduction to Superconductivity}, 2nd ed. (MacGraw-Hill, Inc., New York, 1996). 

\bibitem{bruder} C. Bruder, Phys. Rev. B \textbf{41}, 4017 (1990).

\bibitem{HLS1974} B. I. Halperin, T. C. Lubensky, and S.-K. Ma, Phys. Rev. Lett., {\bf 32}, 292 (1974).

\bibitem{DH1981} C. Dasgupta and B. I. Halperin, Phys. Rev. Lett., {\bf 47}, 1556 (1981).

\bibitem{tesanovic1999} Z. Tesanovic, Phys. Rev. B {\bf 59}, 6449 (1999).

\bibitem{nguyen-sudbo1999} A. K. Nguyen and A. Sudb{\o}, Phys. Rev. B {\bf 60}, 15307 (1999);
A. K. Nguyen and A. Sudb{\o}, Europhys. Lett., {\bf 46}, 780 (1999). 


\end{thebibliography}
\end{document}